\documentstyle[11pt,aaspp4]{article}

\lefthead{Ferrarese \& Ford}  \righthead{Black Hole in NGC 6251}

\def\kms{km s$^{-1}$} \def\an{\AA} \def\sm{M$_{\odot}$}
\def\sl{L$_{\odot}$} \def\dl{$\lambda$} \def\ma{$^{-1}$~}
 \def\ha{H$\alpha$} \def\deg{^{\circ}}
\def\Deg{\hbox{${}^\circ$\llap{.}}}
\def\Min{\hbox{${}^{\prime}$\llap{.}}}
\def\Sec{\hbox{${}^{\prime\prime}$\llap{.}}}

 \def\deg{\hbox{${}^\circ$}}
\def\min{\hbox{${}^{\prime}$}} \def\sec{\hbox{${}^{\prime\prime}$}}

\begin{document}

\title {Nuclear Disks of Gas and Dust in Early Type Galaxies and the
Hunt for Massive Black Holes: Hubble Space Telescope Observations of
NGC 6251}

\author {Laura Ferrarese\altaffilmark{1}} 
\affil{California Institute of Technology, Mail$-$Stop 105-24, Pasadena, CA 91125}  
\altaffiltext{1}{Hubble
Fellow}
 
\author {Holland C. Ford} 
\affil{Johns Hopkins University, 34th and Charles Street, Baltimore MD 21218}

\begin{abstract}

We discuss Hubble Space Telescope optical images and spectra of NGC
6251, a giant E2 galaxy and powerful radio source at a distance of 106
Mpc (for $H_0 = 70$ \kms~Mpc$^{-1}$).  The galaxy is known to host a
very well defined dust disk (O'Neil et al. 1994); the exceptional
resolution of our $V$ and $I$ images allows a detailed study of the
disk structure. Furthermore, narrow band images centered on the
H$\alpha$+[NII] emission lines, reveal the presence of ionized gas in
the inner 0\Sec3 of the disk. We used the HST/Faint Object
Spectrograph with the 0\Sec09 aperture to study the velocity structure
of the disk.  Dynamical models were constructed for two extreme (in
terms of central concentration) analytical representations of the
stellar surface brightness profile, from which the mass density and
corresponding rotational velocity are derived assuming a constant
mass$-$to$-$light ratio $(M/L)_V \sim 8.5$ M$_\odot$/L$_\odot$.  For
both representations of the stellar component, the models show that
the gas is in Keplerian motion around a central mass $\sim 4 - 8
\times 10^8$ \sm, and that the contribution of radial flows to the
velocity field is negligible.

\end{abstract}

\section {Introduction}

Nuclear disks of ionized gas, a few $\times 100$ pc in diameter,
provide an efficient and powerful alternative to stellar dynamical
studies in constraining the central potential of early type galaxies
(Ferrarese, Ford \& Jaffe 1996, Harms et al. 1994, Bower et al. 1998,
van der Marel \& van den Bosch 1998). Since the seminal papers by
Sargent et al. (1978) and Young et al. (1978) dynamical modeling of
the stellar kinematics has been the tool of choice in searching for
massive black holes (BH). However, the method is complicated since the
stellar orbital structure is unknown and difficult to derive from the
observational data (intensity, velocity and velocity dispersion)
because of line$-$of$-$sight projections. More seriously, the well
known degeneracy between velocity anisotropy and varying
mass$-$to$-$light ratio undermines the hope of nailing down the
presence of a massive BH, at least in the brightest, pressure supported
ellipticals (Sargent et al. 1978, Richstone \& Tremaine 1985, van der
Marel 1994).


A new chapter in the search for massive BH was written when the Wide
Field and Planetary Camera on board the Hubble Space Telescope (HST)
was pointed at NGC 4261, a giant elliptical galaxy and powerful radio
source 30 Mpc away. The images showed a regular, flattened disk of
dust and gas, 230 pc in diameter (Jaffe et al. 1993): even if the gas
can easily be perturbed by non gravitational forces (shocks, radiation
pressure, winds, magnetic fields;  Fillmore, Boronson \& Dressler
1986), the regular and flattened structure of the disk implies that
circular motion in the central potential of the galaxy should dominate
the velocity field.  Compared to the stellar system, the disk is
two$-$dimensional, so that complications connected to the projection
of the orbital velocities are not an issue, nor are ambiguities
associated with the presence of velocity anisotropy. Finally, the
bright emission lines from the disk are easy to detect.

These promises were fulfilled when Harms et al. (1994) used the
HST/Faint Object Spectrograph (FOS) to find that the disk of ionized gas
discovered in the Virgo cD galaxy M87 (Ford et al. 1994), is indeed in
Keplerian motion around a $\sim2 \times 10^9$ \sm~central mass, a
result confirmed by later observations with higher spatial resolution
(Macchetto et al. 1997, Ford \& Tsvetanov 1998). The astonishingly
high $(M/L)_V$ of $~3100$ \sm/\sl~within the inner 6 pc lead Ford \& Tsvetanov
to conclude that the central mass must be in the form of a BH.
Spectroscopic data for NGC 4261 (Ferrarese et al.  1996) also point
to a Keplerian velocity field and a $\sim 5 \times 10^8$ \sm~central
mass, likely a BH given the high value $(M/L)_V$ of $2\times10^3$ \sm/\sl~
within 15 pc. Since then, two more BHs have been discovered using HST
kinematics of ionized gas disks.  Bower et al.  (1998) derived a BH
mass of $3 \times 10^8$ \sm~ for NGC 4374, the second brightest radio
galaxy in Virgo. Van der Marel \& van den Bosch (1998) discovered a $3
\times 10^8$ \sm~BH in NGC 7052, an E4 galaxy 59 Mpc away.

This paper discusses HST images and spectra of the giant elliptical
NGC 6251, host of one of the most spectacular radio sources in the sky
(Waggett, Warner, \& Baldwin 1977). At a distance of 106 Mpc (for
H$_0$ = 70 \kms~Mpc$^{-1}$), this is the farthest galaxy for which the
presence of a massive BH has been investigated.  O'Neil et al. (1994)
discovered a small ($\sim$ 1\Sec4 in diameter) disk of dust in the
central region of NGC 6251.  Interestingly, NGC 6251 is one of the
first galaxies (with M87) for which the presence of a nuclear black
hole was claimed, 20 years ago, by Peter Young and his collaborators
(Young et al.  1979), based on the steepening of the surface
brightness profile towards the center. The study of the kinematics of the
ionized gas associated with the dust disk allows us to check 
the black hole claim.

NGC 6251 is an E2 galaxy with m$_B=$13.6 mag (RC3, de Vaucouleurs et
al. 1991) and a  heliocentric velocity $7400 \pm 22$ \kms~(from ZCAT, Huchra et al.
1992).  Optical and UV images have been obtained by Young et al.
(1979), Bender, Doebereiner, \& Moellenhoff (1988), Crane et al.
(1993), and Crane \& Vernet (1997), but it is because of its radio
morphology that the galaxy gained immediate popularity.  Waggett et
al.  (1977) identified it as the parent galaxy of one of the largest
radio sources in the sky: the radio lobes, which show the marked
S$-$shape geometry expected as the result of precession of the feeding
jet, extend for 1\Deg2 in the plane of the sky (over 2 Mpc projected).
The northwest lobe is connected to the small diameter radio core by a
bright, well collimated radio jet extending for 4\Min4 (130 kpc). The
kpc radio structure was studied by Owen \& Laing (1989), Jones et al.
(1986), Willis, Wilson, \& Strom  (1978) and Perley, Bridle, \& Willis
(1984).  
A kpc$-$scale counter$-$jet was seen in the VLA maps by Perley et al.  
(1984), however, a parsec scale counter$-$jet
so far has not been detected with the VLBI (Readhead, Cohen, \& Blandford
1978, Cohen \& Readhead 1979, Jones et al.  1986, Jones \& Wehrle
1994). Jones \& Wehrle (1994) used this observation to conclude that
the jet must lie within 45\deg of our line of sight, based on
Doppler boosting arguments.

X$-$ray observations are discussed by Worrall \& Birkinshaw (1994),
Birkinshaw \& Worrall (1993) and Turner et al. (1997). The gas in the
0.2$-$2.0 keV regime shows a resolved thermal component probably
undergoing a cooling flow, plus an unresolved non$-$thermal component
originating from the inner region of the parsec scale radio jet.

This paper is organized as follows: Section 2 deals with the details
of the HST/WFPC2 data: planning, acquisition,
reduction and a discussion of the images.  The spectra are described in \S 3, 
where line fitting procedures are applied to derive central velocity,
velocity dispersion, line profiles and fluxes from the emission
lines. Different models to the velocity pattern of the emission line
gas are discussed in \S 4, and compared in \S 4.1, where we conclude
that the best fitting models imply the presence of a $\sim 6 \times
10^8$ \sm~central mass, probably a black hole in view of the presence
of an active nucleus. Notes on the accretion mechanisms, and the
origin and morphology of the dust and gas can be found in the
remaining of \S 5. A summary is drawn in \S 6.

\section{HST/WFPC2 Imaging}

\subsection{Observations and Data Reduction}

Narrow band and continuum images of NGC 6251 were obtained on 1996
September 13 with  the HST/Wide Field and Planetary Camera 2 (WFPC2,
Biretta et al. 1996). The F673N filter (central wavelength
\dl$_c$=6732 \an, FWHM $\Delta$\dl= 47 \an) was used to map the
\ha+[NII] emission line region, while the F547M (\dl$_c$=5476 \an,
$\Delta$\dl= 483 \an) and F814W (\dl$_c$=7921 \an, $\Delta$\dl= 1488
\an) filters covered the emission$-$line free continuum on both sides
of the \ha+[NII] emission.

The galaxy nucleus was placed on the 800$\times$800 pixel Planetary
Camera (PC) chip, the highest resolution of the four WFPC2 CCDs: at a
0.046 arcsec/pix scale, each PC pixel covers 24 parsecs at the
distance of NGC 6251. The gain and read out noise of the camera are 7
e$^-$/DN and 7 $e^-$ respectively. The total exposure times of the
observations are 1400 s in F547M, 1000 s in F814W, and 5700 s in
F673N. To facilitate cosmic ray removal, the continuum and narrow band
images were broken into two and four consecutive exposures
respectively.

Routine calibration was performed by the standard HST pipeline and
consisted of correction of small A/D errors, subtraction of a bias
level for each chip, subtraction of a super$-$bias frame, subtraction
of a dark frame, correction for shutter shading effects, and division
by a flat field (see Holtzman et al. 1995a for a detailed discussion).  
Based on the centroids of fifteen point sources present
in the field of view, back to back images taken with the same filter
are registered to within 0.02 pixels. In averaging back to back
images, pixels deviating more than three times the local sigma
calculated from the combined effects of Poisson statistic and read out
noise, were identified as cosmic rays, and replaced with the lowest of
the input values. Images taken with different filters are shifted by
up to one pixel with respect to each other, and were aligned by
performing a bilinear interpolation in both x and y directions.  The
calibrated, combined, and cosmic ray cleaned F547M, F673N and F814W
images are shown in Figure 1, and will be discussed in \S 2.2. The F547M
image is also shown in Figure 2.

Photometric calibration was performed as outlined by Holtzman et al.
(1995b). F547M and F814W counts were transformed to $V$ and $I$
magnitudes respectively, and F673N counts to fluxes. The accuracy of
the photometric zero points is 0.1 mag in $V$ and $I$ and 5\% in the
F673N fluxes.  As further processing, we derived the isophotal
parameters, the reddening map and the pure emission line map as
follows.  Isophotal parameters (position angle, center, ellipticity
and Fourier coefficients) were derived using the IRAF task ELLIPSE,
and are shown in Figure 3. The right ascension, declination and
position angle of the isophotes, averaged between 1\Sec5 and 4\sec,
can be found in Table 1.  The same table lists the coordinates of the
nucleus, calculated as the flux centroid
within a three pixel radius aperture. Figure 4 shows the $V$
brightness profile and $V-I$ color profile, both integrated along the
galaxy isophotes, as a function of the isophotal semi$-$major axis.

A reddening map was derived from the $V$ and $I$ images, in the
assumption that the reddening law follows Cardelli, Clayton and Mathis
(1989), that $R_V \equiv A_V/E(B-V) = 3.1$, and that there is no
discontinuity in the (de$-$reddened) $V-I$ color of the stellar
population across the boundary of the dust region. Figure 4 shows that
the $V-I$ color increases towards the center by about 0.01 mag/arcsec
between 2\sec~and 6\sec. The same trend was assumed within the dusty
regions in calculating the reddening map.  Finally, since the nucleus and
stellar population have different colors,  the nucleus was removed
prior to calculating the reddening map, as described below. The
TinyTim software (Krist 1995) was used to construct theoretical PC
PSFs for each filter and with the appropriate amount of telescope
jitter (jitter information is recorded by the on$-$board telemetry and
was never larger than 7 mas for our observations).  Since the PC PSF
is under$-$sampled (the FWHM is of the order 1.5 pixels), the observed
count distribution is critically dependent on the PSF center within a
pixel.  To avoid sub$-$pixel interpolation in matching the data, the
theoretical TINYTIM PSF was sub$-$sampled by a factor ten, matched to
the data and then re$-$binned. This produces fits accurate to 0.1
pixels.

A contour plot of the $V$ absorption $A_V$, assuming that only half of
the starlight is attenuated by dust, is over$-$plotted on the F547M
image in Figure 5. For a typical dust to gas ratio and dust
properties, the average optical extinction in the disk, $A_V = 0.61
\pm 0.12$ mag, is in excellent agreement with the neutral hydrogen
column density derived by Birkinshaw \& Worrall from ROSAT data.  The
reddening map, adequately scaled to account for differential dust
absorption, was used to de$-$redden the F547M, F673N and F814W
images. The major and minor axes profiles in the central three arcsec
before and after reddening correction are shown in Figure 6.

Finally, a map of the line emitting region was obtained. The $V$ and
$I$ extinction corrected images were averaged and scaled so to match
the median counts of the on$-$band F673N image in a circular annulus
between 1\Sec5 and 4\Sec0 from the nucleus. The resulting continuum
image was then subtracted from the on$-$band image to produce the
emission line map shown in Figure 1.

\subsection{Description of the WFPC2 Images: Dust and Gas Structure}

The most striking feature unveiled by the WFPC2 images shown in Figure
1 is certainly the well defined nuclear dust structure, strongly
reminiscent of the dust disk seen in NGC 4261 (Jaffe et al. 1993,
Ferrarese, Ford \& Jaffe 1996).  The major and minor axis of the
structure extend for 1\Sec43 (730 pc) and 0\Sec34 (170 pc)
respectively: their ratio implies an angle of 76\deg~between the
normal to the disk and the line of sight, assuming the dust is distributed in a thin
circular disk.

As is the case for the NGC 4261 dust disk, the NGC 6251 disk is not
perfectly uniform nor symmetric. This was 
already noted by Crane and Vernet (1997), who further mention the 
possibility that the disk is warped.  The north tip of the disk bends
slightly to the east.  The west side is
brighter than the east side, which is easily interpreted as an
inclination effect: since the line of sight to the far side of the
disk encounters more un$-$obstructed stars, that side will appear
brighter.  Within the errors, the position of the nucleus coincides
with the isophotal center of the galaxy and with the radio core as
measured by Jones \& Wehrle (1994, precessed to J2000), but the disk
is shifted slightly to the north with respect to the stellar
isophotes.  Also, the disk and the galaxy isophotes are not coaxial,
and the disk minor axis is not aligned with the parsec scale radio
jet. 
Quantitatively, the misalignment between the major axes of the disk
and the stellar isophotes is 16\Deg7$\pm$2\Deg4, while the minor axis of the
disk and the radio jet are misaligned by 23\Deg1$\pm$2\Deg4.
This last observation is of particular interest, since it implies
that the jet is not perpendicular to the plane defined by the 
edge of the disk.  If the jet is accelerated and collimated by an
accretion disk threaded by a large scale, perpendicular magnetic field
(e.g. Blandford 1993), then the non perpendicularity between the jet
and the dust disk implies that either the disk is severely
warped, or that there is no continuity between the 100$-$pc scale dust disk
and the sub$-$parsec scale inner accretion disk.  Figure 2 shows the best
fitting elliptical contour to the edge of the disk, and the
directions of the isophotal axes, of the parsec scale radio jet (from
Jones et al. 1986, precessed to the epoch of the WFPC2 observations,
1996.7) and of the acceptable range for the stellar rotation axis (from
Heckman et al. 1985, precessed to 1996.7). All position angles and
coordinates are also summarized in Table 1.

The dust mass in the disk is estimated as $M_{dust}=\Sigma A_B
\Gamma_B^{-1}$ where $\Sigma$ is the surface area of the disk and the
visual mass absorption coefficient is $\Gamma_B = 4 \times 10^4$ mag
cm$^2$ g$^{-1}$ (Sadler \& Gerhard 1985). The mean $A_V$ in the disk,
$0.61 \pm 0.12$ mag, corresponds to $A_B=0.8 \pm 0.2$ mag for
$R_V=3.1$, giving a total dust mass $M_{dust} = (4.1 \pm 0.8) \times
10^4$ \sm.  For a Galactic gas to dust ratio, $M_{gas}/M_{dust} =
1.3\times10^2$, the total mass in the disk is $(5.3 \pm 1.1) \times
10^6$ \sm.

The nuclear emission, probably of non$-$thermal origin, is unresolved
in all images, while the \ha+[NII] emitting region (Figure 1) is
clearly extended.  This is shown in Figures 7a-c, which plot the
radial profiles of the nucleus and the ionized gas region (after
continuum subtraction, \S 2.1), compared to the best matching PSF
radial profile. The reddening$-$uncorrected
magnitudes of the nucleus are $I=17.55 \pm 0.19$ mag and $V = 18.72
\pm 0.14$ mag. To derive reddening corrected magnitudes, we need to add the foreground (galactic) extinction to the mean internal extinction in the disk ($A_V=0.61 \pm 0.12$ mag). According to Schlegel et al. (1998) the foreground reddening in the direction of NGC 6251 is $E(B-V)=0.086 \pm 0.014$ mag (note that this agrees with the Burstein \& Heiles (1984) value of $E(B-V)=0.07$ mag), or $A_V=0.27 \pm 0.04$ mag for $R_V=3.1$. Therefore the total extinction to the nucleus of NGC 6251 is $A_V=0.88\pm0.13$ mag, and the un$-$reddened magnitudes of the nucleus are $V = 17.84 \pm 0.19$ mag and $I = 17.03 \pm 0.20$ mag.
The
\ha+[NII] flux is $(5.8 \pm 0.9) \times 10^{-14}$ erg cm$^{-2}$
s$^{-1}$, corresponding to a line luminosity of $(7.8 \pm 1.2) \times
10^{40}$ erg s$^{-1}$ at a distance of 106 Mpc.  This is comparable to
the line luminosity of the gas disk in NGC 4261 (Ferrarese et
al. 1996). It is also typical of AGNs with similar radio luminosity
(Baum \& Heckman 1989).


\section{HST/FOS Spectra}

\subsection{Observations and Data Reduction}

Thirteen spectra covering the central 0\Sec3 of NGC 6251 were obtained
on 1997 January 11 using the HST/FOS (Keyes
et al.  1995).  The $0\Sec1-$PAIR$-$B square aperture, with projected
dimensions on the sky 0\Sec09 $\times$ 0\Sec09 (corresponding to $46
\times 46$ parsecs at the distance of NGC 6251), was used to map a
3$\times$3 grid, including the nucleus and aligned with the major axis
of the dust disk (position angle 4\deg).  Four additional spectra were
obtained 0\Sec18 on either side of the center of the 3$\times$3 grid,
along the major and minor axes of the dust disk. The position of the
apertures with respect to nucleus are shown in Figure 2. The nucleus
was acquired through a 4$-$stage peak$-$up (Keyes et al.  1995), using
sequentially the 1\Sec0$-$PAIR, 0\Sec5$-$PAIR, 0\Sec25$-$PAIR and
0\Sec1$-$PAIR apertures. While the original intent was to place the
center of the aperture grid on the nucleus, analysis of the map formed
by the peak$-$up steps and dwells shows that the acquisition procedure
was only partially successful, and that the center of the grid is
offset by 0\Sec05 to the north and 0\Sec02 to the west with respect to
the nucleus, due to malfunctioning of the on$-$board acquisition
software.  In the remainder of this paper the aperture positions will
be referred to as labeled in Figure 2.

The grating used, G780H, covers the wavelength region between 6270
\an~and 8500 \an. A spectral resolution of 4.85 $\pm$ 0.02 \an~was
derived from the average FWHM of sixteen emission lines in seven
wavelength calibration spectra taken during the observing
sequence. All spectra are quarter$-$stepped, a procedure in which the
spectrum is stepped onto the linear array of 512 diodes by
one$-$quarter of the diode width (1.43 \an) along the dispersion
direction. All exposure times should have been 2050 seconds; in
practice, to maximize telescope efficiency, positions 3 and E were
exposed for 1950 s, positions N and NUC were exposed for 2060 s and
position W for 2460 s.

The spectra were calibrated by the standard HST pipeline as described
by Keyes et al. (1995). The calibration includes paired$-$pulse
correction, background and scattered light subtractions, and flat
fielding.  In order to achieve a higher accuracy in the wavelength
scale than allowed by the standard processing, we observed arc lamp
spectra either immediately before or after each science spectrum. For
each arc lamp spectrum, dispersion coefficients were computed
following the procedure outlined by Kriss (1994).  The
dispersion coefficients assigned to each science spectrum are the mean
of the coefficients of the two arc spectra immediately preceding and
following it, weighted by the time lapsed between the middle of the
science exposure and the arc spectra.  According to a convention
adopted for the FOS, all wavelengths are vacuum wavelengths.

The largest zero point difference between arc lamp spectra
is 0.45 \an, corresponding to 20 \kms~at 7000 \an.  This defines the
accuracy of our wavelength scale, and is consistent with the FOS
wavelength accuracy limit reported by Keyes et al. (1995).  However,
there is a 2.3 \an~shift between the standard (pipeline processing)
wavelength zero point and the average of the zero points derived from
our arc lamps. This is likely due to the non$-$repeatability in the
position of the filter$-$grating wheel (estimated to be 0.35 diodes,
Keyes et al.  1995).

Inverse sensitivity curves are not available for the 0\Sec1 aperture,
therefore flux calibration is not performed by the pipeline
processing.  Rough fluxes for the 0\Sec1 aperture are obtained by
using the inverse sensitivity curve for the 0\Sec25 FOS square
aperture, multiplied by the ratio of the throughputs of the two
apertures (0.47, Keyes et al. 1995).  Since this ratio is calculated
for point sources, it can be off by as much as a factor $\sim 3.6$
(the ratio of the sky projected areas of the two apertures) in the
extreme case of a source with uniform surface brightness. In addition,
even for point sources, the throughput ratio is strongly dependent
(varying by up to 50\%) on the centering of the source within the
aperture. For these reasons, all absolute fluxes reported in this
paper should be used with extreme caution. However, relative fluxes
are correct to within a few percent.

\subsection {Description of the FOS Spectra: Emission Line Fitting}

The HST/FOS spectra described in \S 3.1 provide an almost complete
coverage of the emitting line region shown in Figure 1.  The nuclear
spectrum is plotted in Figure 8. Emission lines of
[OI]\dl\dl6300,6364, \ha, [NII]\dl\dl 6548,6584, [SII]\dl\dl 6717,6731
and [OII]\dl7325 are detected. As for NGC 4261 (Ferrarese, Ford, \&
Jaffe 1996), which also hosts a very well defined dust disk,  the
[NII]+\ha~complex has broad wings superimposed on narrower lines, and
the [NII] lines are strong compared to \ha. As discussed in more
detail in the next section, the heavy blending of the \ha~and [NII]
lines and the presence of the broad wings make estimating accurate
line fluxes rather difficult. In addition, important diagnostic lines,
such as [OIII]\dl 5007 and H$\beta$, fall outside the wavelength
coverage of the spectra.  However, given the (admittedly not ideal)
[SII]\dl\dl 6717,6731/\ha, [OII]\dl7325/\ha~ and [NII]\dl6584/\ha~line
ratios derived in \S 3.2.1,  the diagnostic plot of Dopita \&
Southerland (1995) show that NGC 6251 lies near the upper limit of the
region defined by LINERS  and well outside the region populated by
Seyfert galaxies, pointing to shocks as the likely cause of ionization
for the gas (Contini 1997, Dopita \& Southerland 1995, Allen, Dopita,
\& Tsvetanov 1998).


The aim of the FOS observations is to obtain a high spatial resolution
map of the velocity field in the central region, to constrain the
central potential. To achieve this, accurate fitting of the emission
lines is necessary. Because of the target mis$-$centering (\S3.1) and the
rapid decrease in the \ha+[NII] surface brightness with radius, we
cannot successfully measure emission lines from the six apertures
which are not adjacent to the nucleus. The [NII]+\ha~complex is
detected at 1.5$\sigma$ at positions 2, 10 and 11, and 2$\sigma$ at
positions 3 and 4, but due to the low S/N and the low resolution of
the spectra, none of the lines can be accurately measured.  The
spectra at the remaining seven aperture positions which will be used
in the dynamical modeling are shown in Figure 9.

\subsubsection{Fits to the The Nuclear Spectrum}

The line synthesis program SPECFIT (Kriss 1994) is used to fit the
spectra. Emission lines are modeled with a Gaussian with four degrees
of freedom: flux, centroid, FWHM and a skew parameter. The continuum
is approximated with a power law, described by two free parameters
(slope and flux at a specified wavelength). Fitting is done
interactively via Chi$-$square minimization using a Marquardt
algorithm.  In the fit, the ratio of the central wavelengths of lines
in doublets ([OI], [NII] and [SII]), is constrained to be equal to the
ratio of the corresponding rest vacuum wavelengths. Also, the FWHMs of
lines in doublets are set equal.  Because of the relative low spectral
resolution of our data, the \ha+[NII] complex is not fully resolved,
and in order to constrain the fit, we fixed the redshift and FWHM of
the \ha~and [NII] lines to be the same.  Because of the limited S/N in
the [OI] and [SII] lines, the shape parameter for all lines is set
equal to the shape parameter of [NII]\dl6584.  Finally, the
[NII]\dl6584/[NII]\dl6548 flux ratio  is fixed to the theoretical
value of three.  These constraints bring the number of free parameters
for the seven emission lines plus the continuum down from 30 to 15.

Figure 10a shows the best fit to the nuclear spectrum obtained under
the assumptions listed above.  The broad wings in the
[NII]+\ha~complex are not reproduced by the model. To obtain a better
fit, we added a broad \ha~component to the model with all
four parameters set free, therefore increasing the total number of
parameters to 19.  The best fit under these conditions is shown in
Figure 10b, and is a clear improvement over our previous
attempt. Quantitatively, the reduced $\chi^2_r$ for the first model
(without broad \ha) is 1.92, while the introduction of the broad
\ha~line brings $\chi^2_r$ down to 1.41.  The parameters of the fit
are listed in Table 2. Within the errors, all narrow lines have the
same redshift, which is however $\sim 100$ \kms~lower than the
systemic velocity of the galaxy, $7400 \pm 22$ \kms~(Huchra et
al. 1992).  The broad \ha~line is very asymmetric, with a steep blue
side and a more gradually declining and more extended red side. In
contrast, no significant asymmetry is seen in the narrow emission
lines. The peak of the broad \ha~line is blueshifted by 870 
\kms~($\Delta\lambda/\lambda=3\times10^{-3}$) with respect to the narrow \ha,
however, because of the marked asymmetry, the midpoint of the line at
half intensity is redshifted by $\sim$850
\kms~from the frame of the
narrow lines.

There are several possibilities for the origin of the broad \ha~
component.  Chen \& Halpern (1989) and Chen, Halpern, \& Filippenko
(1989) studied the emission profiles generated by a relativistic
Keplerian accretion disk around a supermassive black hole. Their
models predict that the line profile will split into a double$-$peaked
structure due to rotation, that the blue peak will be stronger than
the red peak due to relativistic beaming, and that there will be a net
gravitational redshift of the line profile. A sample of 12 radio
galaxies with broad emission lines which can be reproduced in the
Keplerian disk framework is discussed by Eracleous \& Halpern
(1994). The fractional redshift measured for NGC 6251 is comparable to
the mean shift observed for the Eracleous \& Halpern sample, but the
FWHM of the NGC 6251 broad line ($\sim  3700$ \kms) is smaller than
observed for typical disk$-$like emitters ($> 5000$ \kms~with the
distribution peaking at $\sim 12500$ \kms). Alternative mechanisms
leading to double peaked or asymmetric line profiles are biconical or
spherical radial outflows  (Zheng, Sulentic, \& Binette 1990, Zheng,
Veilleux, \& Grandi 1991; see however Livio \& Xu 1997) and massive
binary black holes (Begelman, Blandford, \& Rees, 1980). Higher
spectral resolution and S/N data are necessary to perform detailed
modeling and discriminate between the alternatives mentioned above.

\subsubsection{Fits to the Off$-$Nuclear Spectra}

The fit to the N spectrum (which has comparable S/N to the nuclear
spectrum, and higher than any of the other spectra) was carried out as
described for the nucleus\footnotemark. We found that, while the
central velocity for the narrow lines is significantly different from
the nuclear values (see Table 2), the FWHM, redshift and shape of the
(unconstrained) broad \ha~line are consistent with what measured at
the nuclear location. This confirms the hypothesis that the broad
\ha~line originates from an unresolved region at the nuclear location,
and what is seen in the off$-$nuclear spectra is due to the wings of
the FOS PSF. For all remaining spectra, due to the considerable
decrease in the S/N ratio, it was necessary to constrain the fit by
fixing the central velocity, shape and FWHM (but not the flux) of the
broad \ha~line to the nuclear values. Figure 11 shows the fits to all
seven spectra; the free parameters for each emission line are listed
in Table 2. The fluxes of the broad \ha~emission are found to decline
with increasing distance from the nucleus in a fashion consistent,
within the quoted errors, with what is expected for a point source
centered at the northern edge of the NUC aperture, which coincides
with the nuclear position (\S 3.1).

\footnotetext{No corrections to the fluxes were adopted for
contamination of the off$-$nuclear spectra by light scattered from the
wings of the PSF of the narrow nuclear components, which for a
perfectly centered point source amounts to 12\% and 2\%  of the
nuclear light  in apertures 0\Sec09 and  0\Sec13 away
respectively. These estimates are strongly dependent on the centering
of the point source within the central aperture, and are likely to be
overestimates in our case, since the narrow emission line region is
clearly extended. As was the case for NGC 4261 (Ferrarese et al. 1996)
errors on the central wavelength and FWHM derived from neglecting this
effect are small compared to internal uncertainties on the same
quantities.}

\section {Determination of the Central Mass}

\subsection {The Keplerian Disk Model}

This section will make use of the gas velocities measured at the seven
FOS aperture locations discussed above to probe the presence of a
central mass concentration. We will make the simple assumption
that the velocity field is determined by the combined potentials of
the stars and of the central black hole, and ignore,  for the time
being, the possible presence of radial flows, which will be discussed in
\S4.2. We will further assume
that the gas is confined in a thin disk.

The rotational velocity expected from the stellar potential in which the disk
is embedded can in principle be derived from the observed surface
brightness profile, once a suitable assumption is made for the stellar
mass$-$to$-$light ratio. In practice, the procedure is non$-$unique
since the presence of both the dust disks and the non$-$thermal
nucleus prevents us from performing a complete isophotal analysis 
$-$ and therefore determining the surface brightness profile $-$ in
the critical inner $\sim$ 0\Sec5 (\S 2.1).  Figure 12 shows the best fits to
the  surface brightness profile between 0\Sec5 and 3\Sec5, using two
different analytical functions. The first, represented by the solid
line, is the sum of two exponential functions of the type $I=I_0\times
e^{-r/r_0}$ and, in the assumption of spherical symmetry, can be
deprojected analytically to give the luminosity density plotted in the
lower panel of Figure 12. The second fit to the surface brightness
profile is performed using the Hernquist model (Hernquist 1990). The
two functions give fits of similar quality, however, the extrapolated
central brightness and luminosity density predicted by the Hernquist
model are a factor three and ten higher respectively than given
by the double exponential profile. 

Do the two fits described above give reasonable representations of the 
luminosity density for a galaxy like NGC 6251?
It is now widely recognized that a
well defined relationship exists between the central slope of the
brightness profile   and the absolute magnitude of the host galaxy
(e.g. Ferrarese et al. 1994). In particular, Gebhardt et al. (1996) used a
non$-$parametric approach to determine the logarithmic slope of the
luminosity density dlog$\nu$/dlogr at 0\Sec1 for 42 early type
galaxies observed with the uncorrected HST. At the absolute magnitude
of NGC 6251 ($M_V \sim -22.9$ mag), they find an average slope of
dlog$\nu$/dlogr $\sim -0.6$, while the complete range in slopes
spanned by the observed galaxies  is $-0.3 \leq$ dlog$\nu$/dlogr $\leq
-0.9$.  The double exponential and the Hernquist models applied to NGC
6251, shown in Figure 12, give dlog$\nu$/dlogr $\sim -0.4$ and
$\sim -1.1$ respectively. {\it We can therefore safely assume that the
true stellar mass density, and therefore the mass of the central point
source, are within the limits defined by the two models.} If anything, in fact, 
the Hernquist models produces a slope which is steeper than observed 
for a galaxy of the type of NGC 6251, while marginally smaller slopes than
given by our double exponential fit can be observed. Therefore, if we
err in determining the central luminosity density and mass, we err on
the safe side: our lower estimate on the central mass will indeed
represent a hard lower limit.

Before proceeding, we can provide a further check to the claim that the
real brightness profile is within the limits set by the two fits shown
in Figure 12,  by pushing our measurement of the brightness profile 
closer to the nucleus using the dereddened
images (\S2.1), or the brightness profile along the minor axis of the
dust disk. Note however, that neither method will allow us to bring our
measurements inside the inner 0\Sec2, which are contaminated
by the bright nucleus. The brightness profile along the minor axis of
the dust disk is shown by the open circles and the crosses in Figure
12 \footnotemark, while the brightness profile derived from the dereddened
images is shown by the open triangles. The minor axis
profile agrees well with the double exponential fit, corresponding to
our adopted lower limit on the luminosity density. The dereddened
brightness profile falls in between the two fits, and therefore within
the range of luminosity densities considered in our analysis. Even if
we do not feel confident enough to use these data in constraining the
fits (the position angle and ellipticity of the isophotes are not
known inside 0\Sec5, and assumptions on the intrinsic color of
the galaxy had to be made in de$-$reddening the images), our findings
give us confidence that our two models for the stellar component
indeed include the correct answer.

\footnotetext{The distance along the minor axis of the disk has been
projected to a distance along the semi$-$major axis of the isophotes,
by using at each radius the values of position angle and ellipticity
derived in \S 2.1. Inside 0\Sec5, where the isophotes are not
determined, we assumed a position angle of 25\deg~and an ellipticity
of 0.1, corresponding the the innermost measured values.}

The projected rotational velocity due to a central mass M$_{BH}$ is
given by Harms et al.  (1994) as a function of the angle
$i$ between the normal to the disk and the line of sight, and of the
position angle $\theta$ of the disk major axis. For our analysis, we
introduce two more free parameters, the displacements $\Delta RA$ and
$\Delta DEC$ of the central aperture relative to the center of the
disk.

The expected rotational velocity due to the central mass $M_{BH}$ and
to the stellar potential is calculated at each aperture position, for
both of the stellar models described above. The stellar
mass$-$to$-$light ratio is assumed to be constant and equal to
$(M/L)_V = 8.5$ \sm/\sl~(Faber \& Gallagher 1979). The position angle
$\theta$ and inclination $i$ of the disk are varied between 1\deg~and
90\deg~in 1\deg~increments, and the misplacements $\Delta RA$, $\Delta
DEC$ are sampled in 1/5 of an aperture width (0\Sec018) increments.
For each combination of $i$, $\theta$, $\Delta RA$ and $\Delta DEC$,
the component of rotational velocity due to the stellar potential,
once projected on the plane of the sky, is subtracted from the
measured velocity, and the result, $v$, is compared to the projected
velocity $v_{BH}$ due to a central point source. 

From this point on, we followed the procedure adopted by Ferrarese et
al. (1996).  Combinations of model parameters providing a good fit to
the data will produce a linear correlation between $v_{BH}$
(normalized to $\sqrt{GM_{BH}} = 1$) and $v$; the adopted values of
$\theta$, $i$, $\Delta RA$ and $\Delta DEC$ are then the ones for
which the $\chi^2$ of the least square fit (weighted by the errors on
the observed velocities) is minimized. The solution is calculated
independently for the double exponential and the Hernquist representations of
the stellar component. For the double exponential, the best fitting
model corresponds to $\theta = 62$\deg~$\pm$ 10\deg, $i=32$\deg~$\pm$
15\deg, $\Delta RA = 0\Sec00 \pm 0\Sec018$ and $\Delta DEC = 0\Sec054 \pm
0\Sec018$.  A plot of the predicted velocities $v_{BH}$ versus the observed
velocities (corrected for the velocity component due to the stellar potential 
as described above) is given in Figure 13; the data is indeed very well
fit by a straight line, confirming the assumption of Keplerian motion in the
disk.
As discussed in Ferrarese et al. (1996), the slope and
y$-$axis intercept of the least square fit to the $(v_{BH},v)$ data
points give the central mass and the galaxy systemic velocity
respectively: $M_{BH} = (7.8 \pm 2.3) \times 10^8$ M$_\odot$ and $v_s
=  7370 \pm 13$ km s$^{-1}$. When the Hernquist models is adopted, the
best fit gives $\theta = 57$\deg~$\pm$ 10\deg, $i=39$\deg~$\pm$
15\deg, $\Delta RA = 0\Sec00 \pm 0\Sec018$, $\Delta DEC = 0\Sec054 \pm
0\Sec018$, $M_{BH} = (4.2 \pm 1.4) \times 10^8$ M$_\odot$, and $v_s =
7368 \pm 13$ km s$^{-1}$.  The double exponential and Hernquist models
give fits with comparable $\chi^2$,  corresponding to a 99\%
confidence level, but of course the derived central mass is smaller in
the case of the Hernquist model (which predicts a much higher central stellar luminosity density) than in the double exponential model.
Note that given the errors derived above for the inclination and position angle of
the gas disk, and the errors associated with the subtraction of the 
continuum and nuclear light in producing a pure emission line image (\S 2.1), the
emission line map shown in the lower right panel of Figure 1, and the outline of the gas disk derived from the kinematical analysis (Figure 2) are in full agreement with each other.

The parameters predicted by the Keplerian disk model for the two mass distributions
described above are summarized in Table 3.
The velocity models and the data points are shown in Figures 14 for the
double exponential model and Figure 15 for the Hernquist model. The
velocities have been de$-$projected using the best fitting values for
the inclination and position angle of the disk. In both figures, the
dotted line shows the orbital velocity due to the stellar
potential. For comparison, the dot$-$dashed line is the rotational
velocity due the potential of the disk itself, which is negligible
given the disk small mass, $\sim5 \times 10^6$ \sm. The final
expected velocity, taking into account the black hole, the stars and
the (negligible) contribution of the dust disk, is shown by the thick
solid line.

\subsection {Non$-$Gravitational Motions}

When compared to the stars, the drawback of using ionized gas to
constrain the central potential is that gas is easily pulled around by
non$-$gravitational motions. Especially when only a few velocity data
points are available, the danger of mistaking radial out/inflows for
circular motion needs to be taken seriously. In NGC 6251, a high
degree of gas turbulence is testified by the large line widths, which
are one order of magnitude larger than the (presumably) rotational
velocities, implying that in spite of the fact that the outer dust
disk appears fairly flat, the inner parts must be geometrically thick,
and the possibility of radial flows cannot be excluded.

We consider the simple case in which the gas is moving radially in a
biconical region with velocity $V(r,\psi)=r^a[1-(\psi/\psi_0)^b]$,
where $\psi_0$ is the half opening angle of the cone and $r,\psi$ are
polar coordinates measured relative to the cone apex and the cone axis
respectively.  The exponent $a$ allows us to model both outflows ($a >
0$) and inflows ($a < 0$) with varying acceleration, while the
exponent $b$ controls the angular dependence of the velocity. In
projecting $V(r,\psi)$ on the plane of the sky, the inclination $i$ of
the cone axis with respect to the line of sight, and the position
angle $\theta$ of the projected cone axis, come in as additional
parameters.  The fits are performed using the same procedure as for
the Keplerian disk model.  The parameters providing the best fit are
$a=4.0 \pm 1.0$, $b=5.5 \pm 1.0$, $\psi_0 = 15$\deg~$\pm$ 10\deg,
$i=20$\deg~$\pm$ 15\deg, and $\theta=179$\deg~$\pm$ 10\deg, giving the
velocities plotted in Figure 16 as a function of the observed
velocities.  The biconical flow model and the Keplerian disk model are
discussed and compared in the next section.

\section{Discussion}

\subsection{Comparison of the Velocity Models}

The dynamical models presented in the previous section are based on
seven emission lines. Error$-$bars on the measured velocities are
large, and constraining the model parameters is tricky. However, the Keplerian
disk model does a superior job in fitting the data than the 
biconical flow model. This claim is based on the following observations.

In the biconical flow model, the velocities measured at the W and E
positions are not reproduced in spite of the large inclination and
opening angle of the cone: according to the reduced $\chi^2$, the
model can be rejected at the 40\% confidence level. The inferred
systemic velocity, $v_s=7278 \pm 10$ \kms, provides a bad match to the
ZCAT value of $7400 \pm 22$ \kms; finally, the position angle of the
cone axis is misaligned by $\sim 60$\deg~with respect to the axis of
the radio jet, implying that the radio jet and the ionized gas are not
collimated by the same mechanism. We conclude that  outflows (or
inflows), if present, do not dominate the velocity field in the inner
0\Sec2.

For both stellar models, the Keplerian disk model reproduces the data
at the 99\% confidence level. It predicts a systemic velocity, $v_s
\sim  7369 \pm 13$, in complete agreement with the ZCAT value. It
places the center of the gas disk at the exact position of the flux
centroid determined in a completely independent manner (cf. \S
3.1). In fact, while formally we have used a six parameter fit
to our seven velocity data points, three of those parameters
are well determined from independent measurements, and have been
left free to vary as a consistency check.
It appears as if the only failed prediction of the model is the
misalignment, by $\sim 30$\deg, between the minor axis of the disk and
the radio jet, and even this might not be a serious problem. The disk
of ionized gas predicted by the model is seen more face on (by $\sim
40$\deg) than the larger dust disk, and the position angles of the two
disks are misaligned by $\sim 55$\deg.  This implies that the entire
structure is tilted and warped, and it is therefore not unjustified to
assume that the warp persists inwards and that at the parsec$-$size
scales, which are inaccessible even at HST resolution, the disk indeed
becomes perpendicular to the parsec scale radio jet. Possible causes
for the warping of the disk will be discussed later \S5.3.

As discussed in the previous section, the two stellar models adopted
to reproduce the kinematics of the disk of ionized gas, encompass the
range of stellar mass density expected for a galaxy with the
luminosity of NGC 6251. Hence, the inferred masses for the central
point source, $M_{BH} = (7.8 \pm 2.3)\times 10^8$ \sm~and  $M_{BH} = (4.2 \pm 1.4)
\times 10^8$ \sm, represent an upper limit and a hard lower limit
respectively to the true central mass.  The mass to light ratio in the
inner 46 pc (corresponding to the FOV of the FOS aperture) is $(M/L)_V
\sim 20$ \sm/\sl~if the Hernquist model is adopted, and
$\sim 160$ \sm/\sl~for the double exponential model.  We conclude that the
kinematical data imply the presence of a central mass concentration,
$4-8\times 10^8$ \sm, supporting the central hypothesis of the AGN
paradigm, according to which the nuclear activity is powered by a
massive black hole. Radial flows play, if anything, only a
secondary role. Note also that, since the kinetic energy trapped in the turbulent
motion of the gas has not been accounted for, any number derived here for the
central mass is likely an underestimate of the true value.

\subsection{The Accretion Mechanism}

Advection dominated accretion flows (ADAF, Narayan \& Yi 1995,
Abramowicz et al.  1995, Narayan, Yi, \& Mahadevan 1995) have received
considerable attention in the past few years as a plausible mechanism
to reconcile the observational limits on black hole masses and
accretion rates in low luminosity AGNs, with the surprisingly small
amount of energy radiated by the active nucleus (Fabian \& Canizares
1988, Reynolds et al. 1996, Lasota et al. 1996, Di Matteo \& Fabian
1997). At very low accretion rates ($< \alpha^2 \dot M_{Edd}$, where
$\dot M_{Edd}$ is the accretion rate corresponding to the Eddington
luminosity, and $\alpha$ is the Shakura$-$Sunyaev viscosity parameter),
ADAF results in a radiative efficiency $10^2 - 10^3$ times smaller than
expected from a standard thin accretion disk (Di Matteo \& Fabian
1997). This is a consequence of the fact that in ADAF the cooling
timescale for the ions exceeds the inflow timescale, and most of the
thermal energy is advected inwards rather than being radiated locally.
ADAF have been successfully advocated to reproduce the nuclear
luminosity and spectral energy distribution of M87, NGC 4258 and M60
(Reynolds et al. 1996, Lasota et al. 1996, Di Matteo \& Fabian 1997)


Is ADAF necessary to explain the nuclear non$-$thermal luminosity
observed for NGC 6251, in the light of the newly determined estimate
of the central mass?  ROSAT/PSPC observations (Birkinshaw \& Worrall
1993) place lower limits on the central pressure and proton density of
the ISM of $P_0 = 2\times 10^{-11}$ N m$^{-2}$ and $n_{p0} = 0.1$
cm$^{-3}$ respectively.  Assuming normal cosmic composition ($n_e/n_p$
= 1.18), these limits give a Bondi (1952) accretion rate of $\sim 5 \times  10^{-4}$ \sm~yr$^{-1}$. Converted at 10\% radiation efficiency,
the corresponding accretion luminosity is $\sim 3 \times 10^{42}$
for a $\sim 6 \times 10^8$ \sm~black hole\footnotemark.

\footnotetext{The value quoted is likely a lower limit to the Bondi's
accretion rates, since lower limits to the pressure and density of the
ISM have been used. If the physical parameters of the ISM in M87 are
taken as characteristics, then the Bondi's accretion rate and accretion
luminosity for NGC 6251 would be a factor five higher than quoted in
the text. This does not change the qualitative nature of our
arguments.}

Radio, optical and X$-$ray measurements of the nuclear non$-$thermal luminosity are
plotted in Figure 17.  The radio data are from VLBI observations by
Jones et al. (1986) with 20 mas resolution, the optical data are as
derived in \S 2.1 ($\sim$ 0\Sec1 resolution), while the X$-$ray
data are ROSAT/PSPC measurements by Birkinshaw \& Worrall (1993)\footnotemark~ with
25\sec~resolution, and ASCA data from Turner et al. (1997) between 2 keV and
10 keV. \footnotetext{The ROSAT data include only the unresolved component at 1 keV
corresponding to model 4a of Birkinshaw and Worrall}
The spectral energy distribution of M87 (which can
only be explained in a ADAF scenario) is also shown for comparison
(Reynolds et al. 1996). The non$-$thermal nuclear flux is 30 times brighter in NGC 6251 than in M87 at every wavelength. A rough lower limit to
the total  luminosity of NGC 6251 is calculated assuming spectral
indexes ($L \propto \nu^{-\alpha}$) $\alpha=-0.25$, 0.52, 1.0, and 1.26
in the radio, optical, 0.5$-$2 keV and 2 keV$-$10 keV X$-$ray ranges respectively
\footnotemark, and integrating the resulting spectral
energy distribution for $\nu < 10$ keV (the upper frequency of the
ASCA observations). The result, $L \sim 3 \times 10^{43}$ erg s$^{-1}$,
is compatible with Bondi's lower limit, and therefore accretion through
a standard, geometrically thin, accretion disk is consistent with the
observations.

\footnotetext{The radio and X$-$ray spectral indexes are from Jones et
al. (1986), Birkinshaw \& Worrall (1993) and Turner et al. (1997),
while the optical spectral index is a simple fit to the $V$ and $I$
de$-$reddened fluxes for the nucleus given in \S 2.1}

ADAF might still be taking place, but would require an accretion rate
$10^2-10^3$ larger than the Bondi's value, in order to account for the
observed luminosity. Indeed, given the difficulty encountered by
standard accretion disk models in reproducing the hard X$-$ray
component (Krolik 1998), ADAF might be necessary to account for the
observed ratio of optical to X$-$ray luminosity (Lasota et al. 1996).
The accretion rate required for NGC 6251 in an ADAF scenario, $0.1-1$
\sm~yr$^{-1}$, is compatible with the 0.3 \sm~yr$^{-1}$ mass deposition
rate estimated by Birkinshaw \& Worrall (1993) from the cooling flow of
the thermally unstable, extended, hot ISM.




\subsection{The Warping of the Disk and the Origin of the Dust}

In \S 5.1, we concluded that the dust/gas structure seen in the nuclear
region of NGC 6251 is warped: the angle between the normal to the disk and the line
of sight varies from $\sim 35$\deg~in the inner region to $\sim
76$\deg~in the outer parts, and the position angle of the major axis
changes by $\sim 55$\deg. We will discuss this observation in the
framework of three different scenarios.

\subsubsection{Triaxiality} 

Evidence for triaxiality is difficult to obtain. Even if not direct
proof, the small isophotal twist (Figure 3), the slight misalignment of
the radio jet and the isophotal minor axis (Figure 2) and the fact that
NGC 6251 is not rotationally supported ($v/\sigma =0.16$, Heckman et
al. 1985) are consistent with the presence of some degree of
triaxiality. Since the kinematical signature of triaxiality is 
a misalignemt between the apparent rotation axis and minor axis of the galaxy, high
S/N, high dispersion spectra taken along the apparent major and minor axis
of the galaxy would provide additional information as to the existing level
of triaxiality. 
Note that the presence of a massive black hole is expected
to destroy triaxiality (Merritt \& Quinlan 1998), but the process would
proceed on timescales longer than a Hubble time for a giant elliptical
such as NGC 6251.

The settling time of the gas in an axisymmetric
galaxy is of the order $10^8 - 10^9$ yr (Tohline, Simonson, \& Caldwell
1982). Since under the assumption of axisymmetry the equilibrium
configuration must lie on the symmetry plane perpendicular to the
rotation axis, the misalignment between the principal axes of the galaxy
and the axes of the dust disk, and the warping of the disk are
difficult to explain if the origin of the gas in internal, and can be
explained in a merging scenario only if the gas has been acquired
within the past few $\times 10^8$ years.

However, in a triaxial galaxy which is slowly tumbling around one of
its principal axis (as could be the case for NGC 6251, which  has a
rotational velocity of $47 \pm 16$ \kms, Heckman et al. 1985), four
dynamically different stable solutions are possible (van Albada,
Kotanyi, \& Schwarzschild 1982, Steiman$-$Cameron \& Durisen 1984). When
the dust and gas occupy the anomalous orbits which have an axis roughly
perpendicular to the system's rotation axis, a change in orientation of
the anomalous plane with radius leads to warping of the dust structure. 
The sense and appearance of the warp depends on the orientation
of the tumble axis, tumble rate, axis ratio, and radial mass
distribution of the galaxy.

A detailed modeling of the rotation figure of NGC 6251 is beyond the
scope of this paper, but we note that triaxial tumbling systems have
been invoked to reproduce the warping of the dust disks in NGC 5128 and
M84 (van Albada et al. 1982), and might therefore be responsible for
the peculiar geometry of the NGC 6251 dust disk.

\subsubsection{Merging}

An external origin for the dust in early type galaxies is generally
advocated (e.g. Goudfrooij et al. 1994, van Dokkum \& Franx 1995).
A search of the CfA ZCAT redshift catalog (Huchra et al. 1992) shows
that NGC 6251 is in a low density environment. The nearest cluster, ZW
1609+8212, is over 8 Mpc away in projected distance and has several
thousand \kms~higher redshift. Within $\sim 0.7$ Mpc (the median
pairwise radius of loose groups, Ramella, Geller, \& Huchra 1989,
Zabludoff \& Mulchaey 1998) three objects are found, none of which
with known redshift. However,
the luminosity in the resolved thermal X$-$ray gas in NGC 6251 is
$\sim 9 \times 10^{41}$ erg s$^{-1}$ in the 0.5$-$2.5 keV range
(Birkinshaw \& Worrall 1993, Worrall \& Birkinshaw 1994) typical of
loose groups (Mulchaey \& Zabludoff 1998), but is concentrated in a
region (3\min~$\sim 90$ kpc) much smaller than the typical median
pairwise radius of loose groups. The possible implication is that NGC
6251 is the evolutionary end$-$point of a loose group, in which all
galaxies have coalesced to form a giant elliptical (Toomre 1977; Weil,
\& Hernquist 1994). 

Whether or not the above scenario has in fact taken place, there is
one major difficulty with the hypothesis that the dust and gas in NGC
6251 has been assimilated from dwarf systems.  In NGC 6251, and any
other LINER, the [NII] emission is always stronger than \ha. Even the
most extreme (in terms of strength of the magnetic field and shock
velocity) shock ionization models of Dopita \& Southerland (1995)
cannot reproduce [NII]/\ha~ flux ratios higher than $\sim$1.5,
assuming solar abundances for the gas. For NGC 6251, [NII]/\ha~ is
$\sim 3-4$, implying that the gas metallicity is at least solar, and
possibly a factor a few higher.  On the other hand, the typical
metallicity range spanned by dwarf systems is between 1/10 to 1/100
solar (Caldwell et al. 1992, Lisenfeld \& Ferrara 1998); it is
therefore unlikely that dwarf systems are progenitors of the gas seen
in NGC 6251.

\subsubsection{Radiation driven warping}

The arguments in \S 5.3.1 and \S 5.3.2 seem to imply that an external 
origin for the dust is unlikely, and therefore the warping of the
dust/gas disk could correspond to a stable configuration in a triaxial
system. We now consider a further possibility.  An accretion disk which
is illuminated by a central radiation source will become unstable to
warping, provided that the disk is optically thick to both absorption
and re$-$emission of radiation (Pringle 1996, Maloney, Begelman, \&
Pringle 1996, Livio \& Pringle 1997).  If the disk is optically thick,
the radiation absorbed at a particular point in the disk will be
re$-$emitted by the same point perpendicularly to the disk surface. If
the initial configuration of the disk is perturbed, so that the disk is
illuminated in a non$-$uniform manner, the back$-$reaction of
radiation  will produce a net torque and induce warping instabilities
in the disk, which will start to wobble or precess. This mechanism
is successful in explaining the point$-$symmetric
morphology observed in some planetary nebulae (Livio \& Pringle
1997),  the warping of the disk of water masers in NGC 4258 (Maloney
et al. 1996), and of the $\beta$ Pictoris disk (Armitage \& Pringle
1997).

The time scale for the instabilities to set in is given by Livio \&
Pringle as a function of the mass of the disk, and the radius, mass and
luminosity of the central object. By taking the values in \S 2.2 for the
radius and mass of the disk, and replacing the Schwarzschild radius as
the radius of the central object, the timescale for the instabilities
to set in is $t_w \sim 5 \times 10^9$ yr. The calculation assumes a
$6 \times 10^8$ \sm~black hole, radiating $\sim 3 \times 10^{43}$ erg
s$^{-1}$. Our assumption are more likely to lead to an
underestimate than an overestimate of $t_w$, since the disk is not
optically thick at all wavelengths ($A_V = 0.6$), and the radiation
field is not necessarily isotropic. Radiation induced warping is
therefore not playing a significant role in shaping the disk.

However, if the disk becomes optically thin at the boundary between the
inner gas disk and the outer dust disk, then $t_w$ would drop by the
ratio of the masses of the two disks, or about a factor $10^3$. In this
case, radiation driven warping would affect the inner, ionized gas
dominated, part of the disk, while the outer part would be left
unperturbed. Interestingly, the value of $t_w$ in this case, a few
$\times 10^6$ yr, is of the same order of magnitude as the precession
period $t \sim 1.8 \times 10^6$ yr quoted by Jones et al.  (1986) to
explain the S$-$shaped symmetry of the large scale radio lobes. This
would be expected if the jet is emerging perpendicular to the disk, and
therefore changes its direction on timescales $\sim t_w$ as the disk
precesses.

In conclusion, radiation warping might be responsible for the twisting
between the inner and outer parts of the disk, by acting on the
optically thick ionized gas disk. It would not, however, influence the
shape and orientation of the outer dust disk.


\section{Conclusions}

This paper discusses HST optical images and spectra of NGC 6251, a
giant E2 galaxy and powerful radio source at a distance of 106 Mpc
(for $H_0 = 70$ \kms~Mpc$^{-1}$). The $V$ and $I$ WFPC2 images, which
have 0\Sec1 (51 pc) resolution, show a well defined dust disk, 730 pc
in diameter, whose normal is inclined by 76\deg~
to the line of sight. A significant ionized gas component is confined
to the central $\sim 0\Sec3$ of the disk. The FOS 0\Sec09 square
aperture was used to map the velocity of the gas in the central
0\Sec2. Dynamical models have been constructed assuming two different
analytical representations for the stellar brightness profile,
designed to encompass the acceptable range  of central luminosity
density for a galaxy of the type and luminosity of NGC 6251.  For
both representations of the stellar mass density (and therefore stellar
potential), the kinematics of the gas imply the presence of a central
mass concentration, $4 \times 10^8$ to $8 \times 10^8$ \sm, likely a 
black hole given the presence of nuclear activity. The
contribution of radial flows to the velocity field is negligible. 
The bolometric non$-$thermal luminosity of the nucleus is higher than
the lower theoretical constraint given by accretion at the Bondi rate
and 10\% efficiency (appropriate for a conventional thin accretion
disk).  Therefore, advection dominated flows do not necessarely have
to be advocated to account for the energy output, given the newly
determined black hole mass estimate.

The dynamical modeling also shows that the inner parts of the disk,
dominated by the ionized gas, are significantly twisted with respect
to the outer, dusty regions. Neither are perpendicular to the radio
jet; however, the perpendicularity at the parsec scale remains to be
tested.  Even if the X$-$ray properties of NGC 6251 make it a possible
candidate for the evolutionary end$-$point of the coalescence of a
loose group,  it is unlikely that the metal rich gas in NGC 6251 could
have been  assimilated from metal poor dwarf galaxies, and an internal
origin for the dust and gas  seems more likely. Cooling flows have been
advocated (Birkinshaw \& Worrall 1993) and a stable warped
configuration for the dust/gas disk is possible in the presence of a
tumbling triaxial potential. We also considered the role of radiation
driven warping in shaping the disk, and concluded that it has a
negligible effect, at least in the outer, dust dominated, regions.

The models presented in this paper are based on seven emission lines.
Error$-$bars on the measured velocities are large, and constraining the
model parameters is tricky. A total of fourteen orbits of HST time were
spent in obtaining the NGC 6251 spectra, therefore higher quality data
would be prohibitively expensive. At a distance of $\sim 100$ Mpc, NGC
6251 represents the limit to which the search for massive black holes
could be pushed with the HST/FOS: a second trip deep inside the NGC
6251 gas disk with HST/STIS will undoubtedly enlighten the picture.

\acknowledgments

We are pleased to thank Dr Mario Livio for useful comments.
LF acknowledges support by NASA through Hubble Fellowship grant
HF-01081.01-96A   awarded by the Space Telescope Science Institute,
which is operated by the Association of Universities for Research in
Astronomy, Inc., for NASA under contract NAS 5-26555
The work  presented in this paper  is  based on observations  with the
NASA/ESA Hubble  Space  Telescope,  obtained by  the   Space Telescope
Science Institute, which is operated by AURA, Inc. under NASA contract
No. 5-26555.  Support  for this work   was provided by NASA  through
grant GO-06653.01-95A from STScI.

\clearpage

\begin{figure}
\figurenum{1}
\plotone{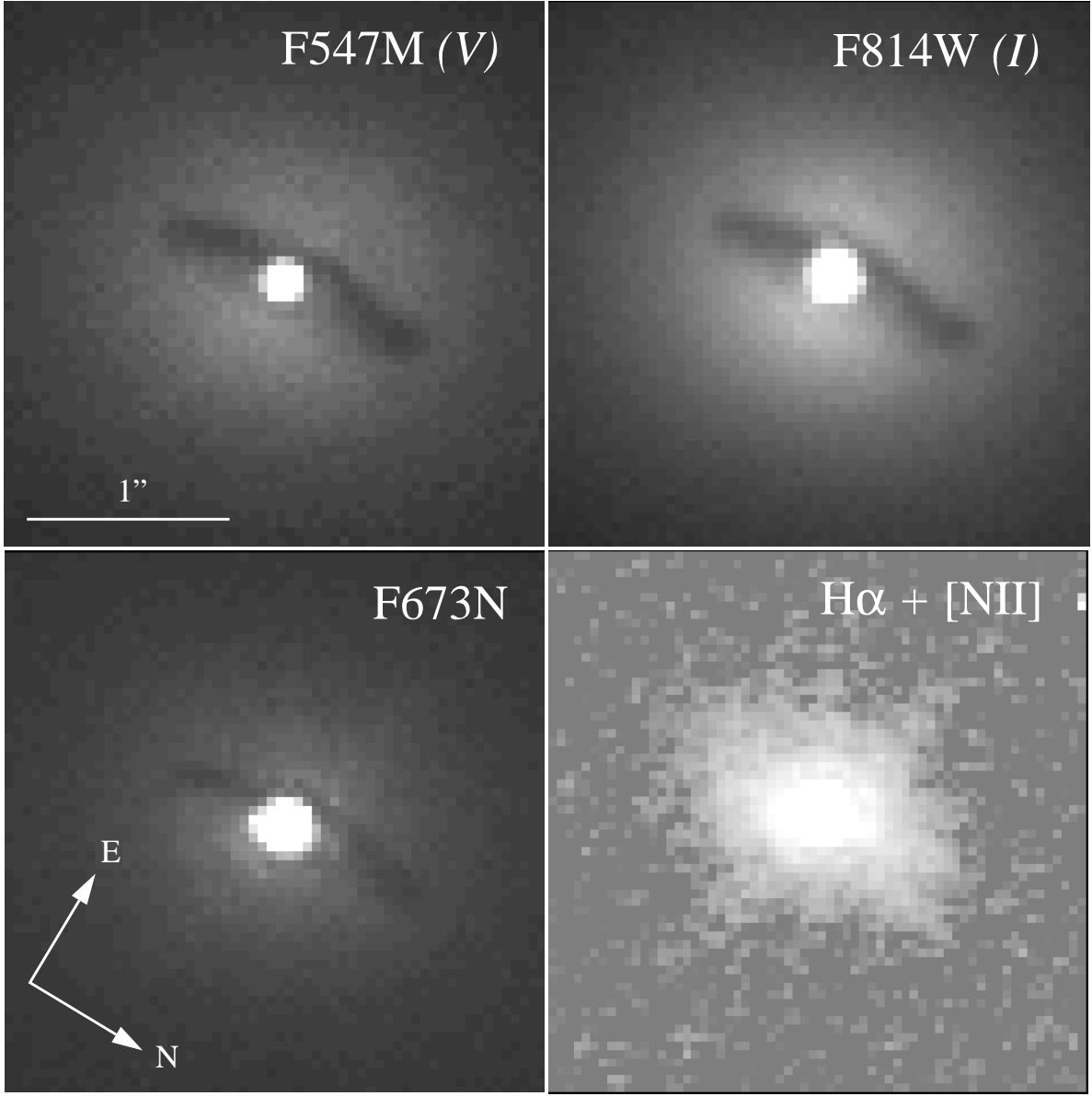}
\caption{HST/PC images of the central 2\Sec7$\times$2\Sec7 of
NGC 6251, in the continuum F547M and F814W bandpasses and in the
\ha+[NII] on$-$band F673N filter. The lower right panel shows the
continuum subtracted and de$-$reddened emission line image, derived as
described in \S 2.1}
\end{figure}

\clearpage

\begin{figure}
\figurenum{2}
\plotone{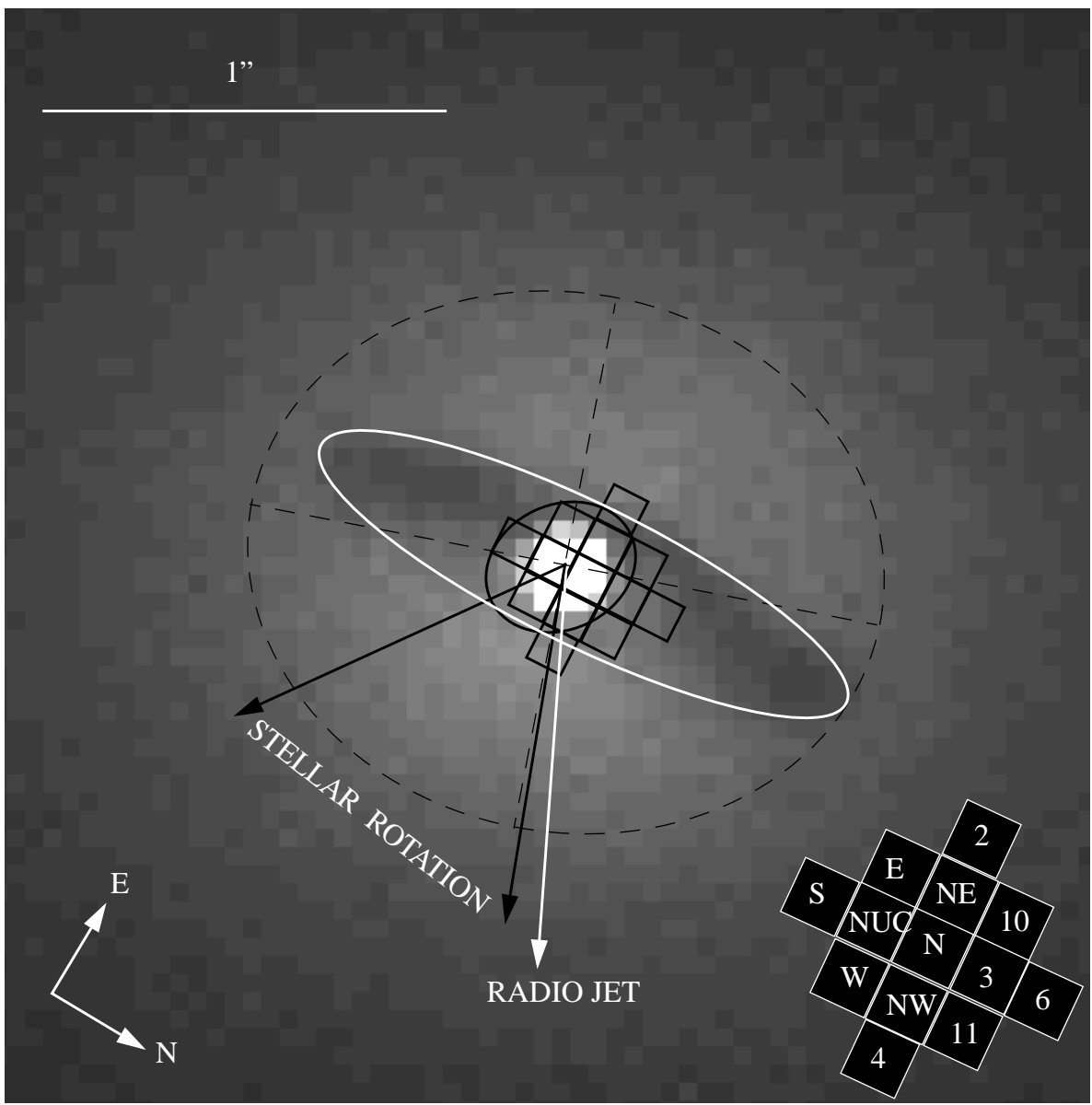}
\caption{The inner 2\Sec7$\times$2\Sec7 of the F547M PC image of
NGC 6251. The larger dashed ellipse is the best fitting elliptical isophote,
while the large solid white contour is the best fitting ellipse to the
edge of the dust disk. The smaller black ellipse represents the 
emission line disk, derived from the kinematic
model as described in \S 4.1. The position angle of the radio jet is
from Jones et al. (1986), while the acceptable range for the stellar rotation
axis is from Heckman et al. (1985).}
\end{figure}

\clearpage

\begin{figure}
\figurenum{3}
\plotone{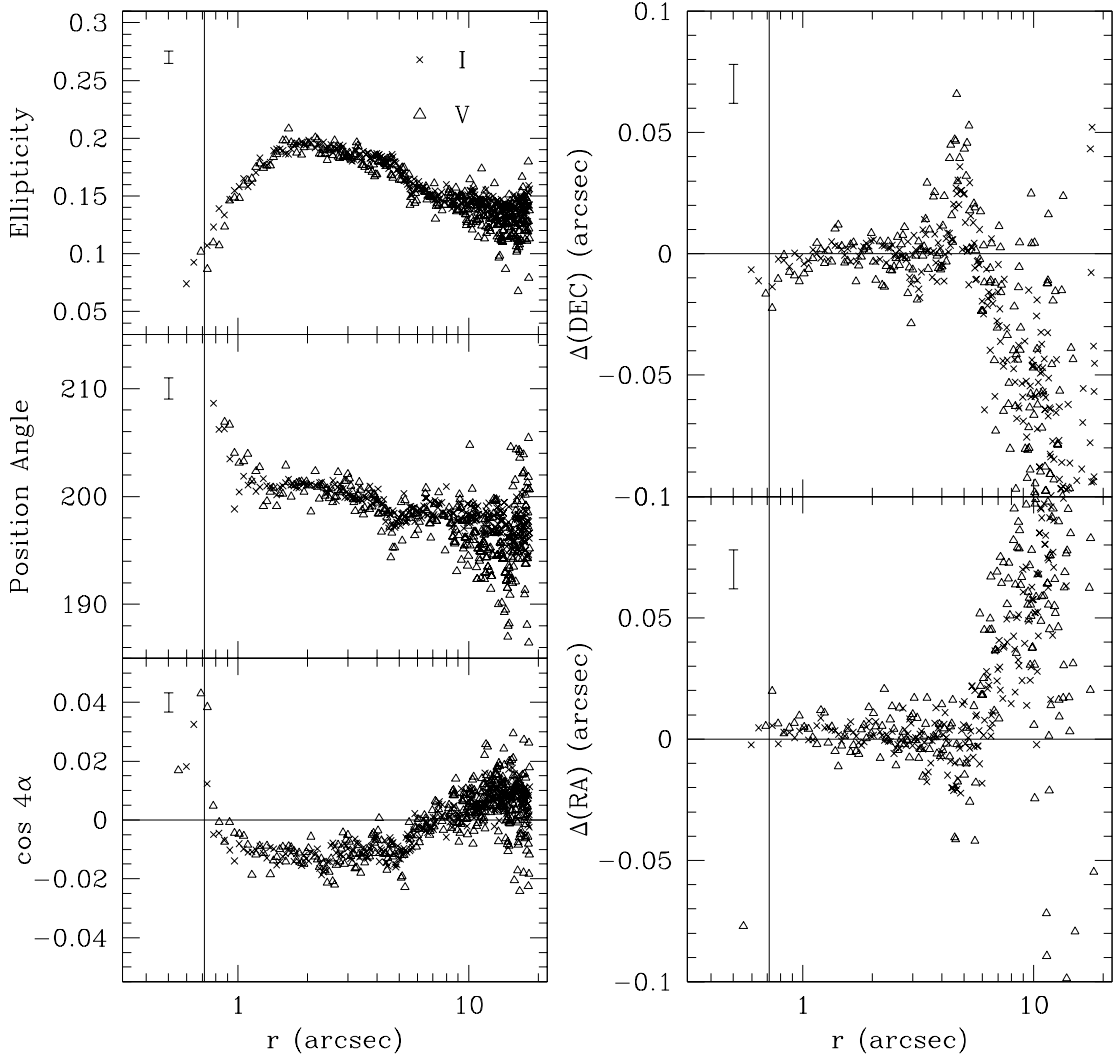}
\caption{Ellipticity, major axis position angle, fourth order cosine
Fourier coefficient and isophotal center as a function of semi major
axis length in F547M (triangles) and F814W (crosses). The solid
vertical line marks the edge of the dust disk. The zero point in
RA and DEC is defined as the mean isophotal center between 1\Sec5 and
4\sec (Table 1). Typical errorbars are shown in the Figure.} 
\end{figure}

\clearpage

\begin{figure}
\figurenum{4}
\plotone{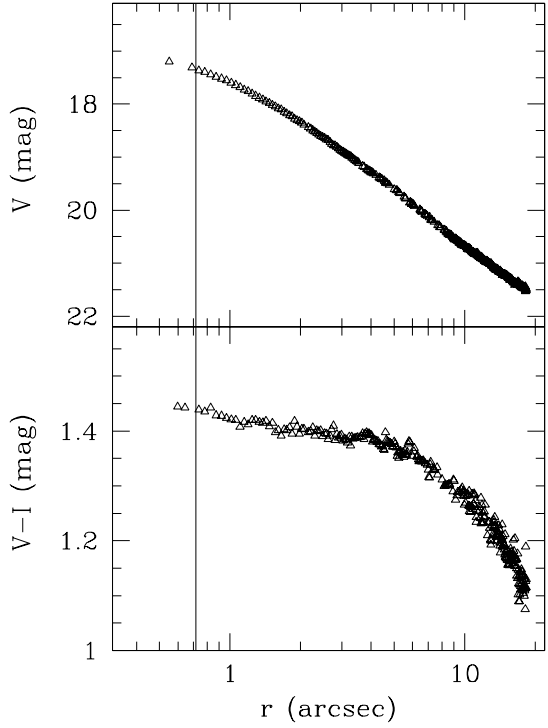}
\caption{The $V$ brightness profile and the $V-I$ color, both integrated along the isophotes, as a function of the semi$-$major axis length. The solid
vertical line marks the edge of the dust disk. Note that
the magnitudes plotted are not corrected for either foreground ($E(B-V)=0.086$ mag) nor (in the inner region) internal extinction (see \S 2.2 for further details). The nucleus, which has $V-I$ = $1.2 \pm 0.2$ (or $(V-I)_0 = 0.8 \pm 0.3$ mag corrected for both internal and foreground extinction),
is therefore bluer than the stellar component, as expected.}
\end{figure}

\clearpage

\begin{figure}
\figurenum{5}
\plotone{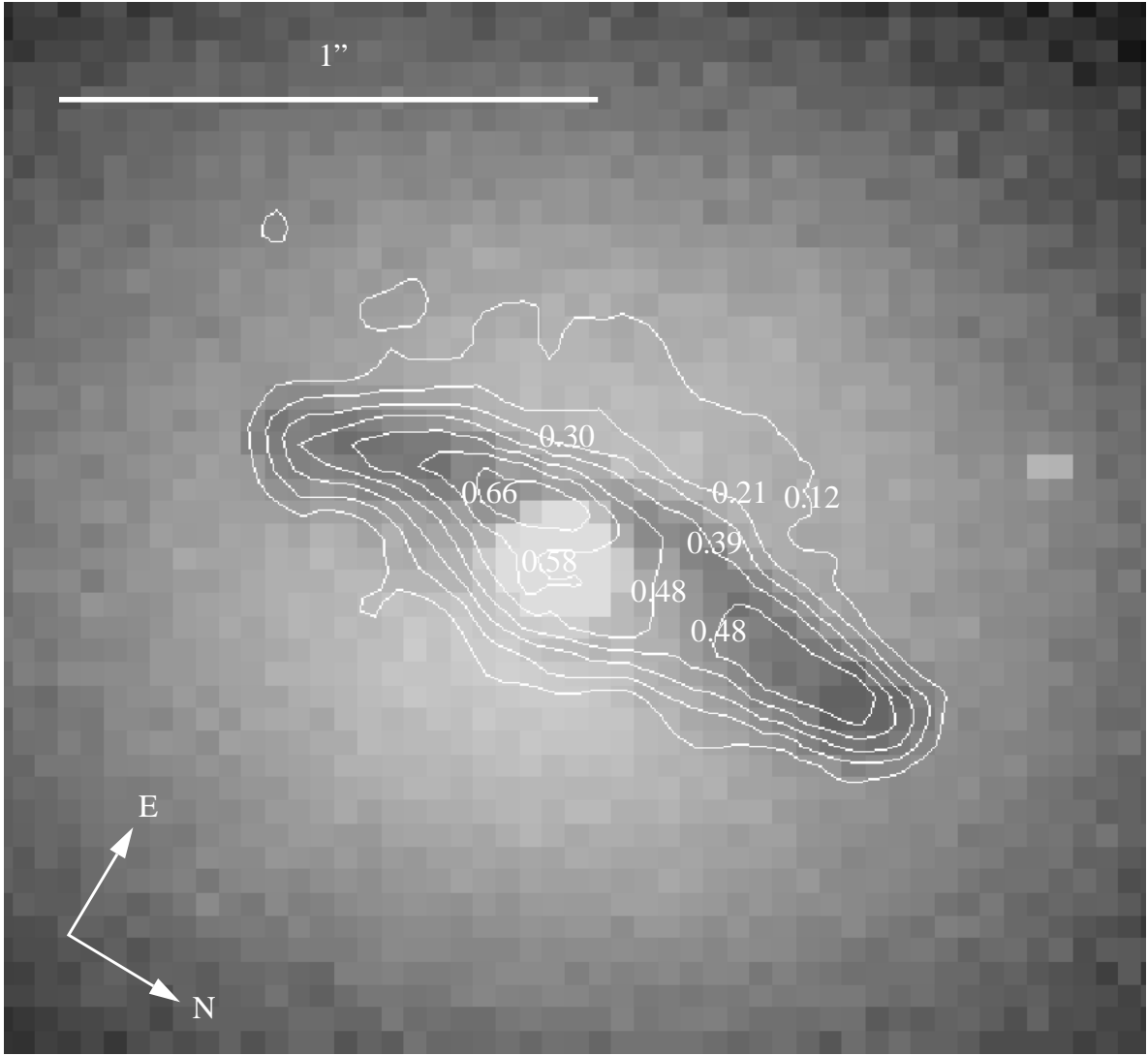}
\caption{The reddening map superimposed to the F547M image of the dust disk. The contours are labeled by the values of $A_V$ (mag). The image size is $2\Sec3 \times 2\Sec3$.}
\end{figure}

\clearpage

\begin{figure}
\figurenum{6}
\plotone{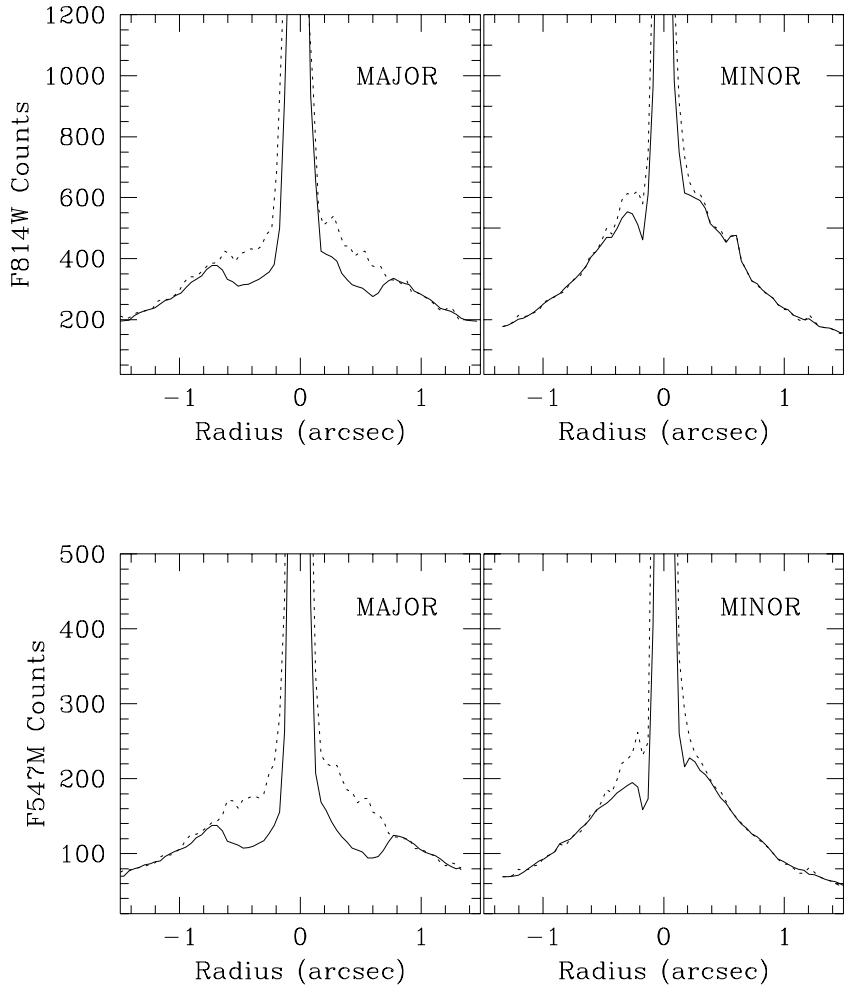}
\caption{Cuts through the nucleus along the major and minor axes of the dust disk, for both F547M and F814W, before (solid line) and after (dashed line) reddening correction.}
\end{figure}

\clearpage

\begin{figure}
\figurenum{7a}
\plotone{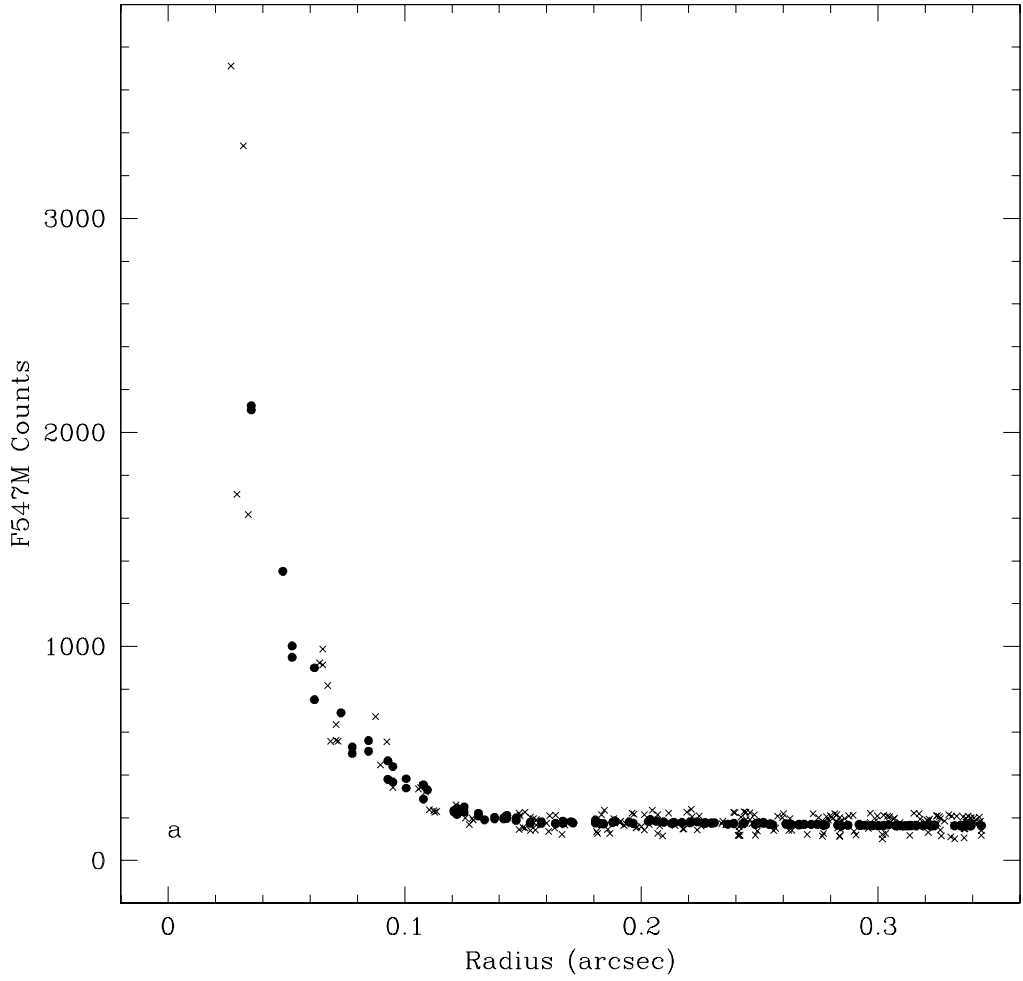}
\caption{A radial plot of the nucleus in F547M (crosses), with
over$-$plotted the best fitting scaled radial profile of a theoretical PSF (dots).}
\end{figure}

\clearpage

\begin{figure}
\figurenum{7b}
\plotone{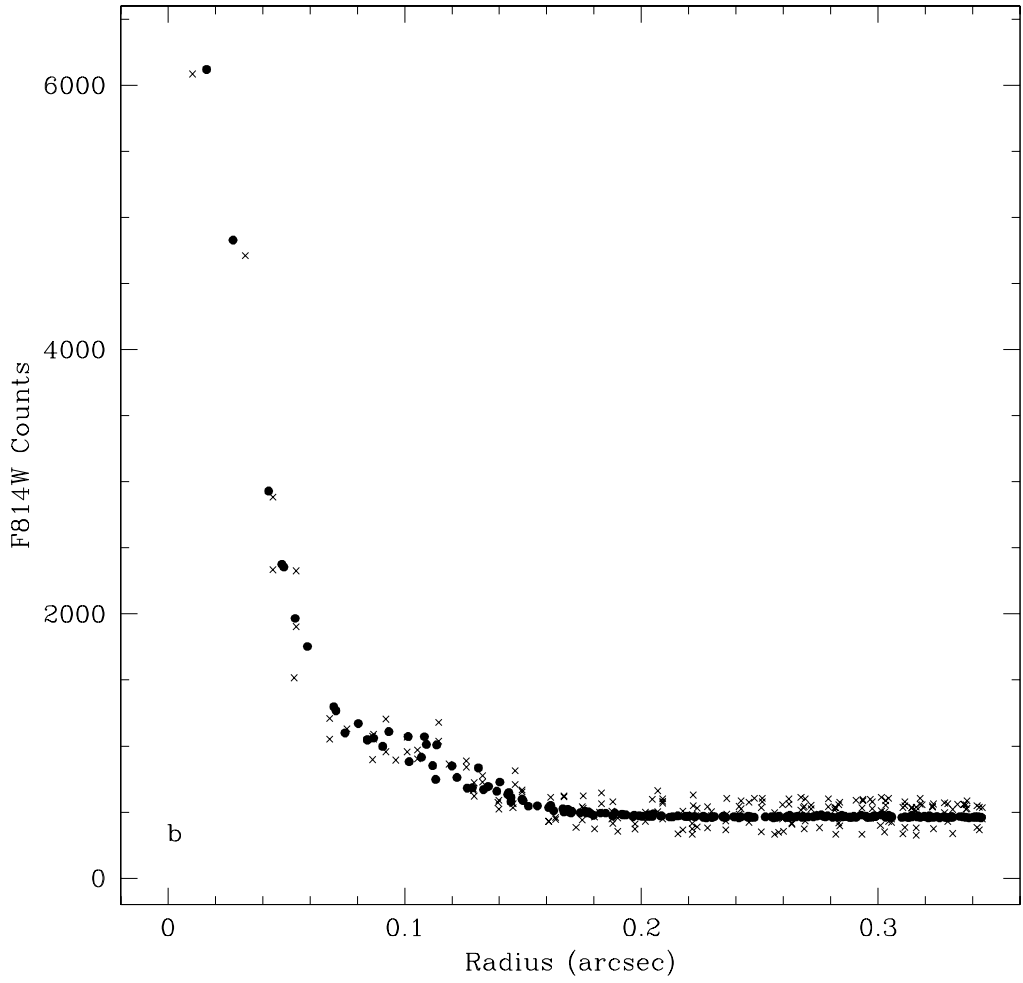}
\caption{As in Figure 7a, but for F814W.}
\end{figure}

\clearpage

\begin{figure}
\figurenum{7c}
\plotone{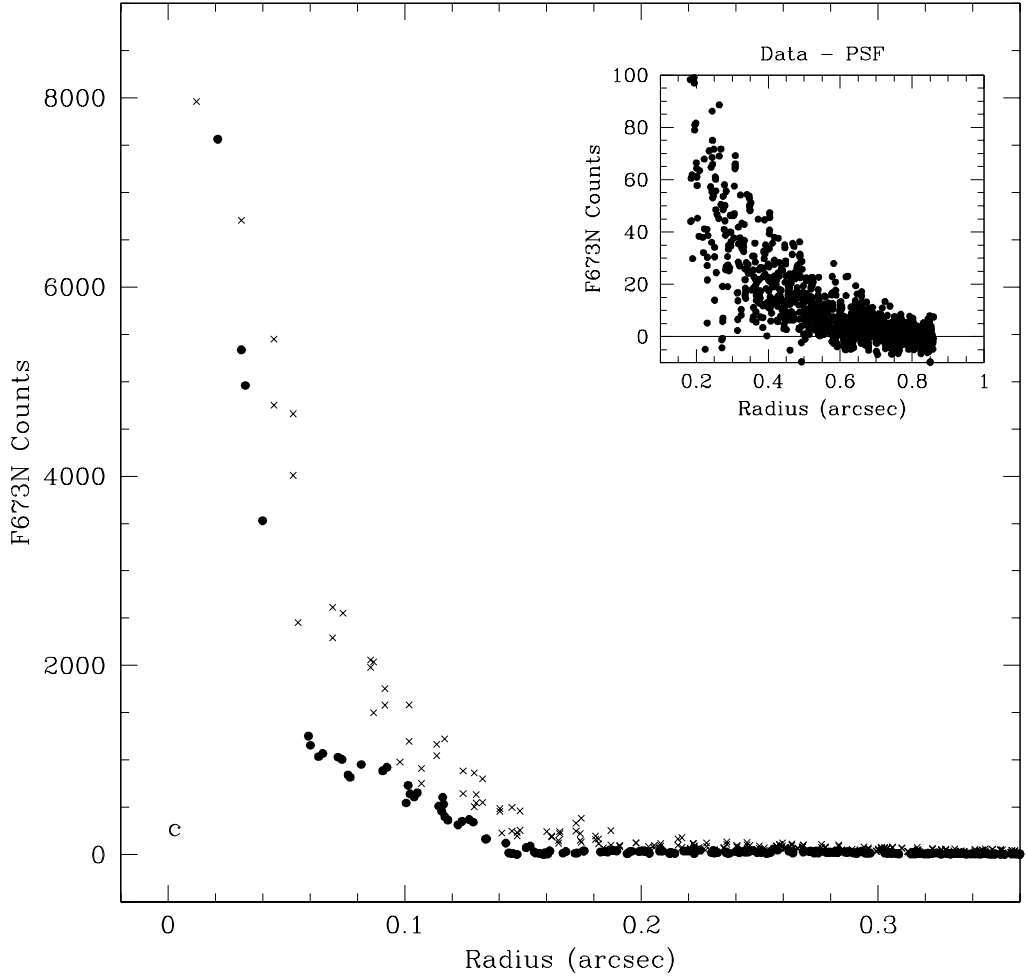}
\caption{As for Figure 7a, but for the \ha+[NII] emission line region (\S 2.2).
The inset shows the residual of the PSF fit in the inner arcsec, clearly
indicating that the emission is resolved.}
\end{figure}

\clearpage

\begin{figure}
\figurenum{8}
\plotone{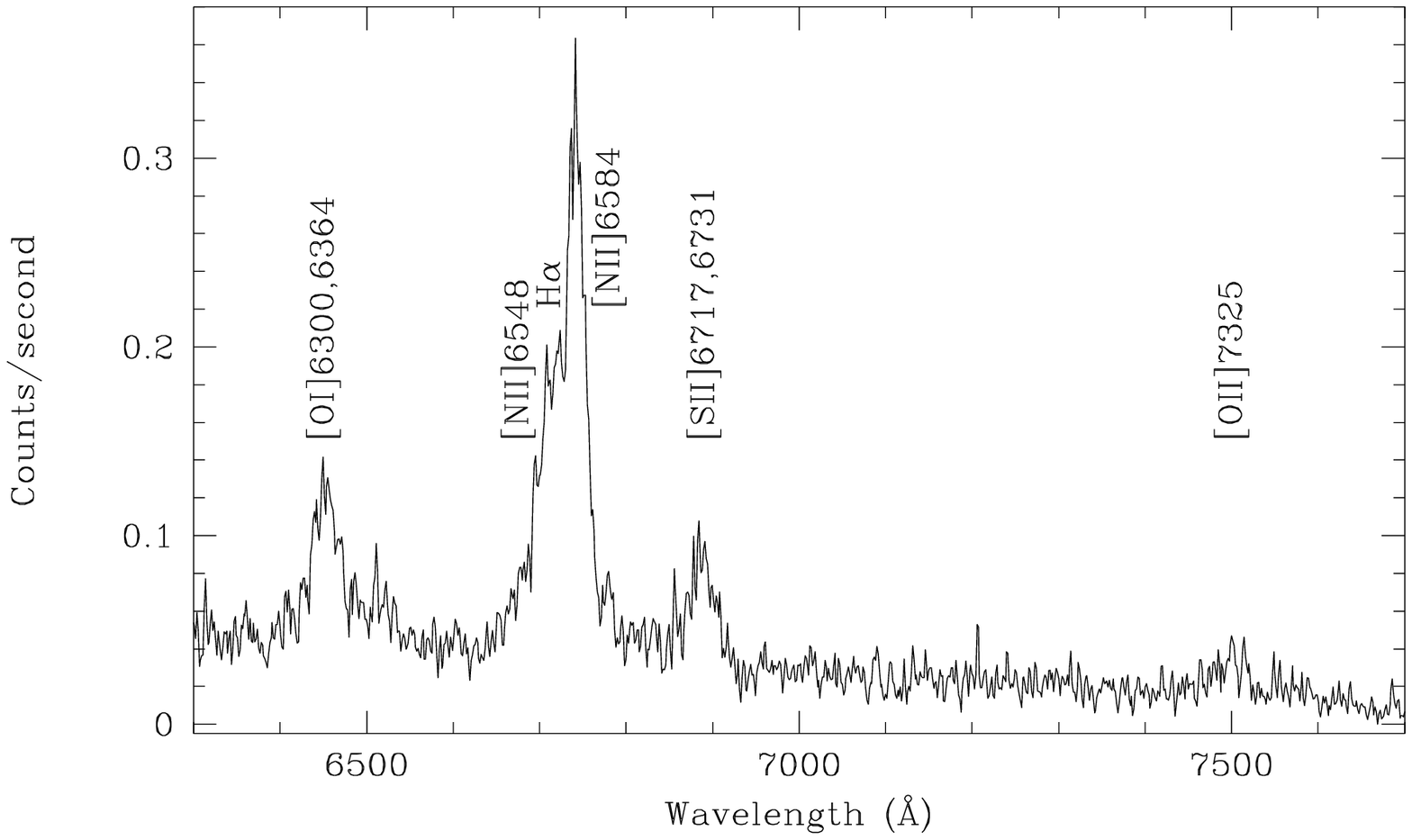}
\caption{A detail of the HST/FOS spectrum at the nuclear position (NUC, Figure 2).}
\end{figure}

\clearpage

\begin{figure}
\figurenum{9}
\plotone{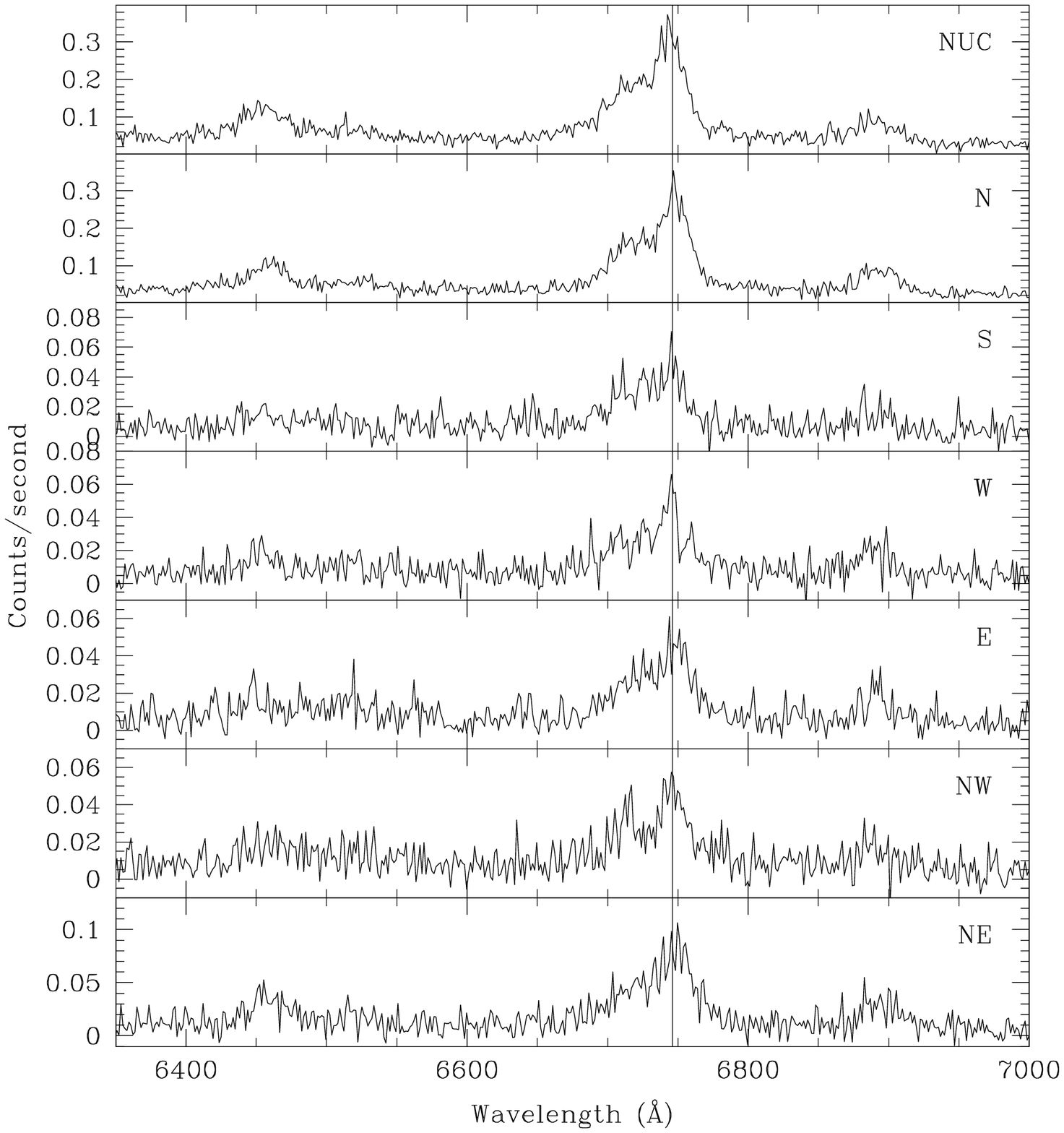}
\caption{A composite picture showing spectra at the seven locations at
which emission lines are detected. The vertical line marks
the position of \ha~at the systemic
velocity of the galaxy (7400 \kms).}
\end{figure}

\clearpage

\begin{figure}
\figurenum{10a}
\plotone{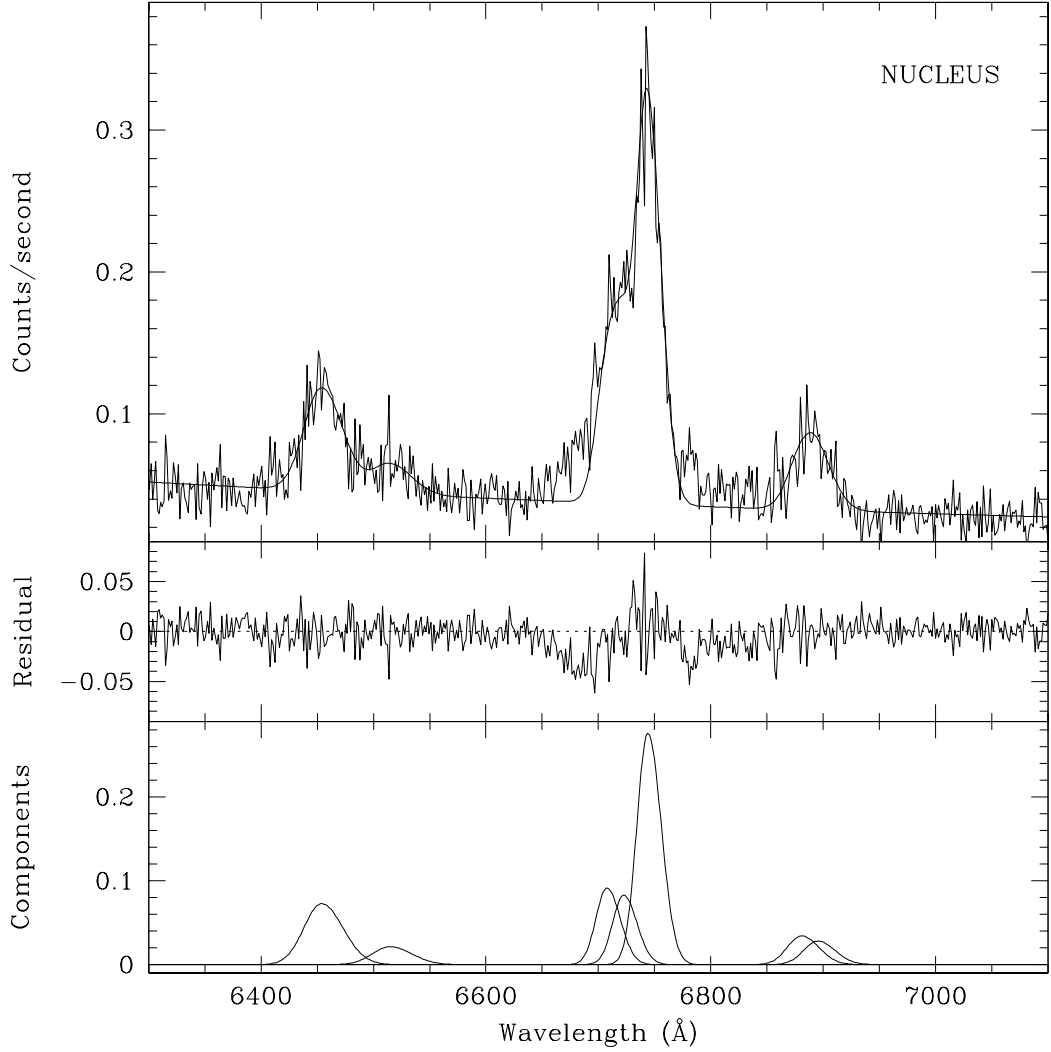}
\caption{Best fit to the nuclear spectrum obtained by assuming that
only one velocity component is present for each emission line. The
broad wings of the \ha+[NII] complex are not well fit by the model.}
\end{figure}

\clearpage

\begin{figure}
\figurenum{10b}
\plotone{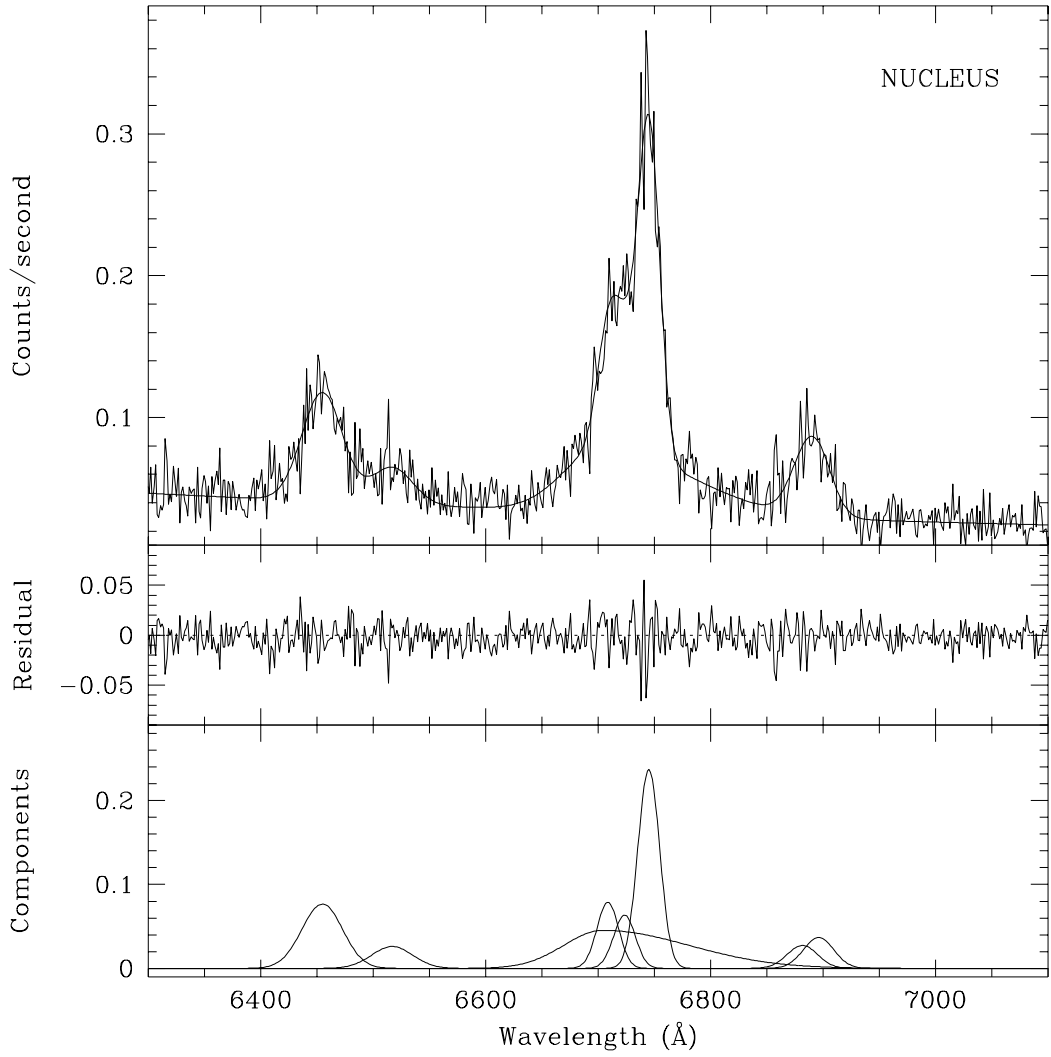}
\caption{Best fit to the nuclear spectrum obtained by assuming that the \ha~emission
originates in two distinct regions, one giving rise to a very broad
component, the other coincident with the [NII] emitting region. The
broad wings of the \ha+[NII] complex are now well fit by the model }
\end{figure}

\clearpage

\begin{figure}
\figurenum{11}
\plotone{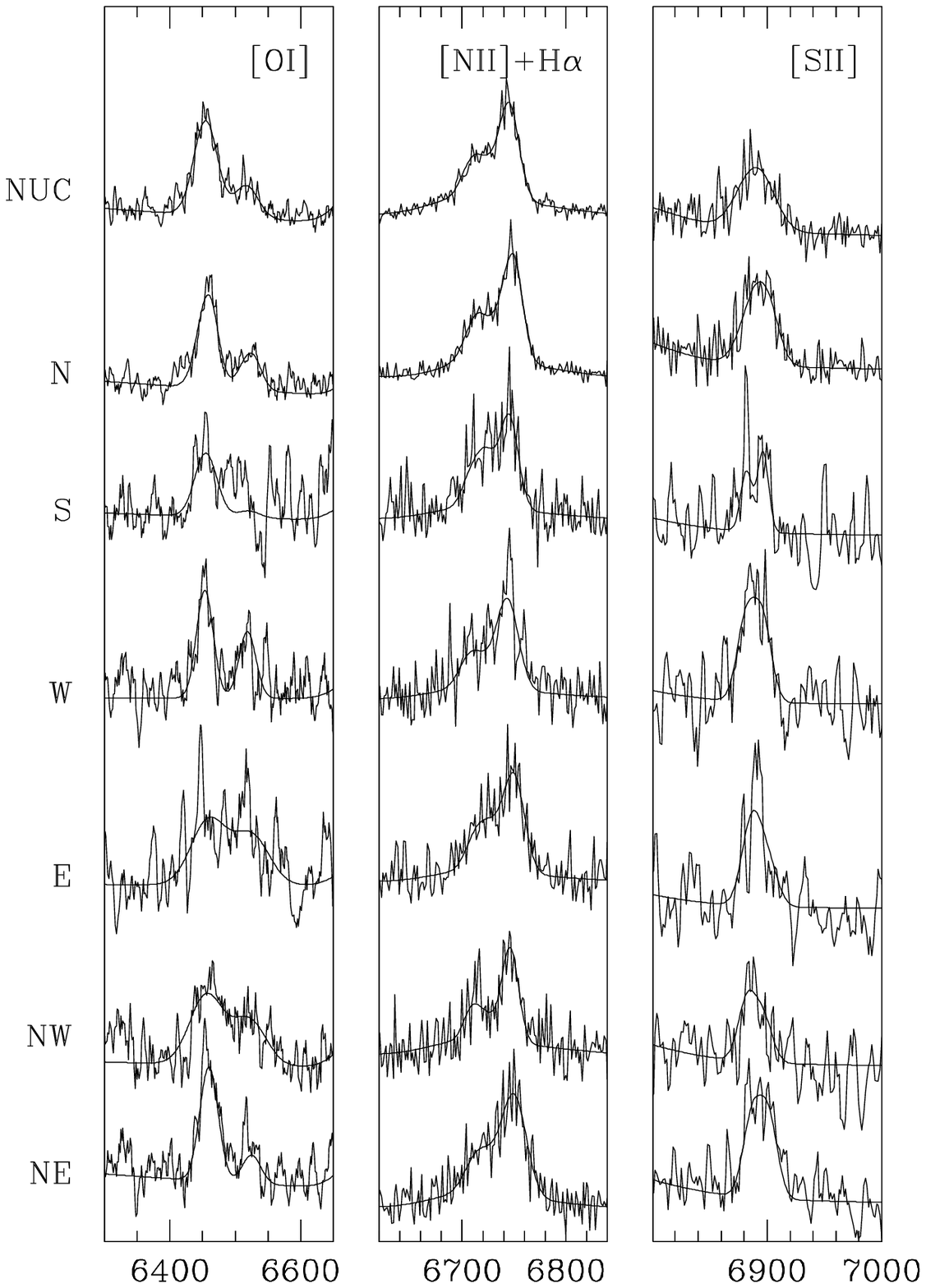}
\caption{Best fits to the [OI], [NII]+\ha~and [SII] lines obtained as
described in \S 3.2.2 for all spectra}
\end{figure}

\clearpage

\begin{figure}
\figurenum{12}
\plotone{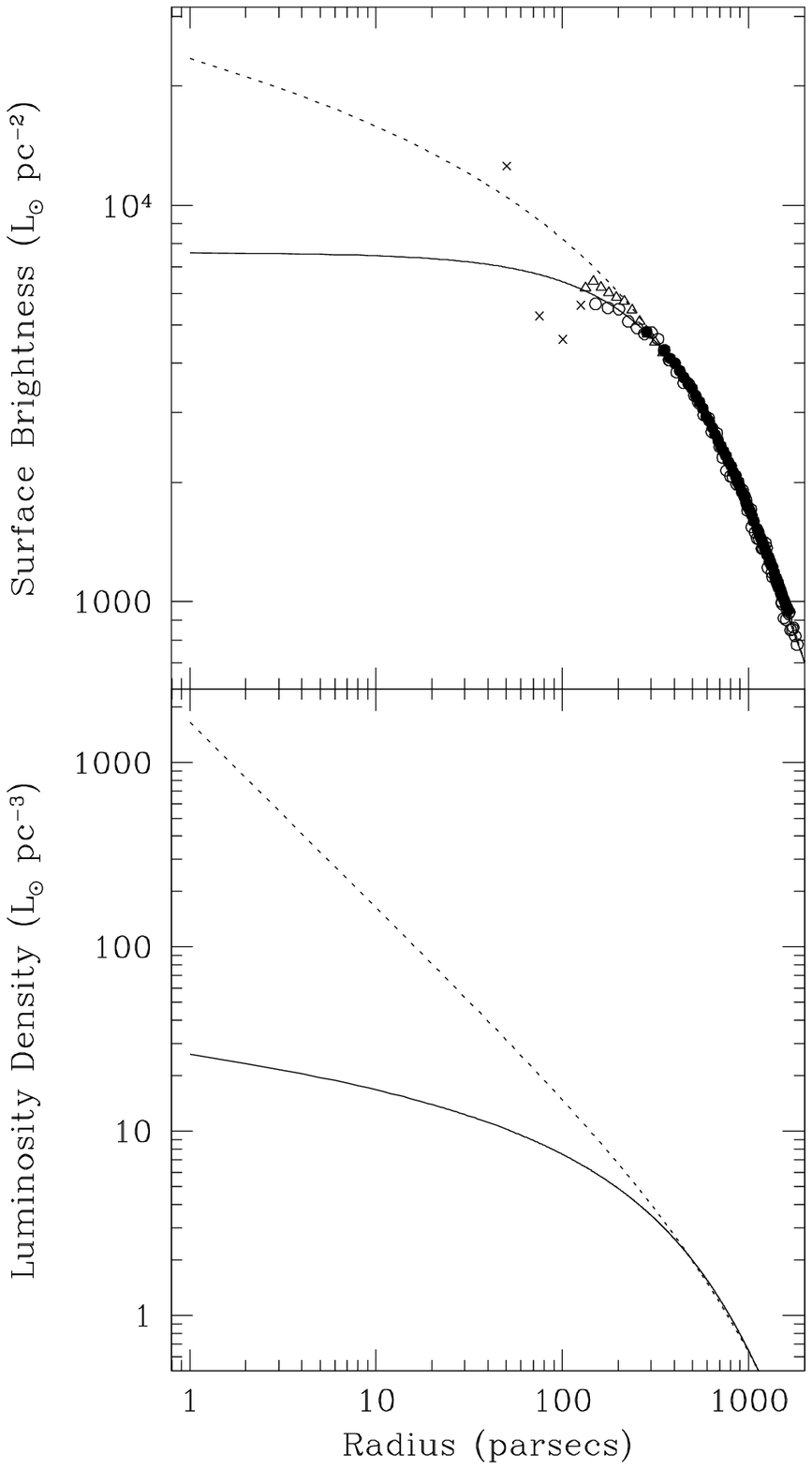}
\caption{The upper panel shows the best fits to the observed $V$
surface brightness profile (integrated over the isophotes, and shown
by the filled circles, see \S 2.1) between 0\Sec5 and 3\Sec5 using a
double exponential profile (solid line) and a Hernquist model
(dashed line). The brightness profile determined along the minor axis
of the dust disk (projected along the major isophotal axis as
described in \S 4.1) is shown by the open circles and the crosses: the
crosses are   used to show data points affected by the presence of the
dust disk or the bright nucleus. The open triangles show the
brightness profile determined from the dereddened images.
The lower panel of
the figure shows the deprojected luminosity density corresponding to
the double exponential and Hernquist models. The deprojection was done
in the assumption of spherical symmetry.}
\end{figure}

\clearpage

\begin{figure}
\figurenum{13}
\plotone{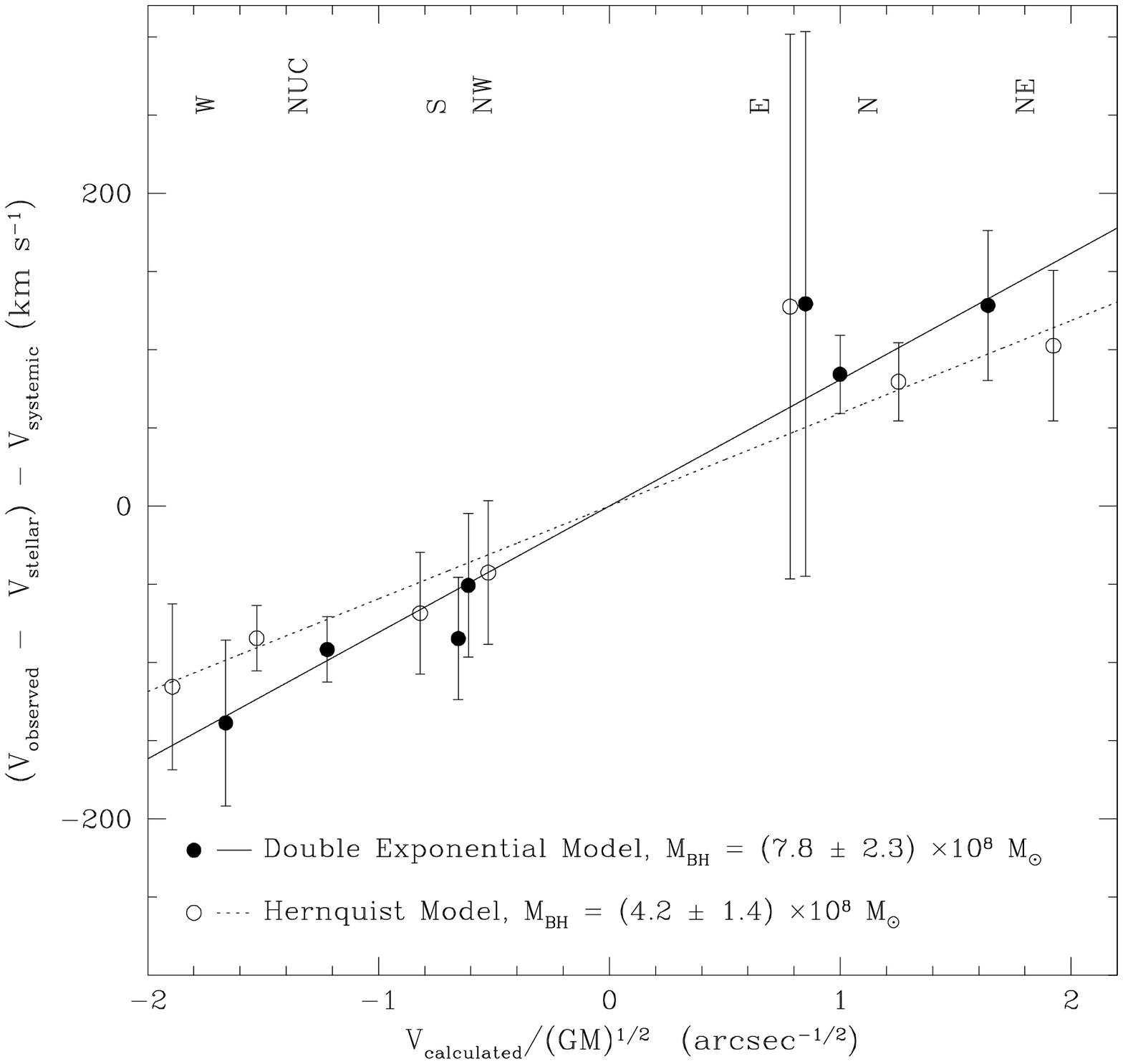}
\caption{Predicted velocities vs. observed velocities for the best
fiting Keplerian model. The rotational velocity component due to the
stellar potential has been subtracted from the observed velocities,
assuming a double exponential surface brighthess profile (solid
circles) or a steeper Hernquist model for the luminosity density (open
circles). In both cases, we assumed a mass to light ratio for the
stars of 8.5 /sm\/sl. The systemic velocity of the galaxy has also
been subtracted from the observed velocities. The predicted velocities
are normalized to (GM$_{BH}$)=1, and are in units of arcsec$^{-1/2}$.
The best least square fit to the data points when the double
exponential model is adopted, is given by the solid line, and yields a
black hole mass (proportional to the slope of the fitted line) of
$(7.8 \pm 2.3) \times10^8$ \sm. The dotted line gives the best fit to
the data points when the Hernquist representation of the stellar
component is adopted, and yields a black hole mass of $(4.2 \pm 1.4)
\times10^8$ \sm.}
\end{figure}

\clearpage

\begin{figure}
\figurenum{14}
\plotone{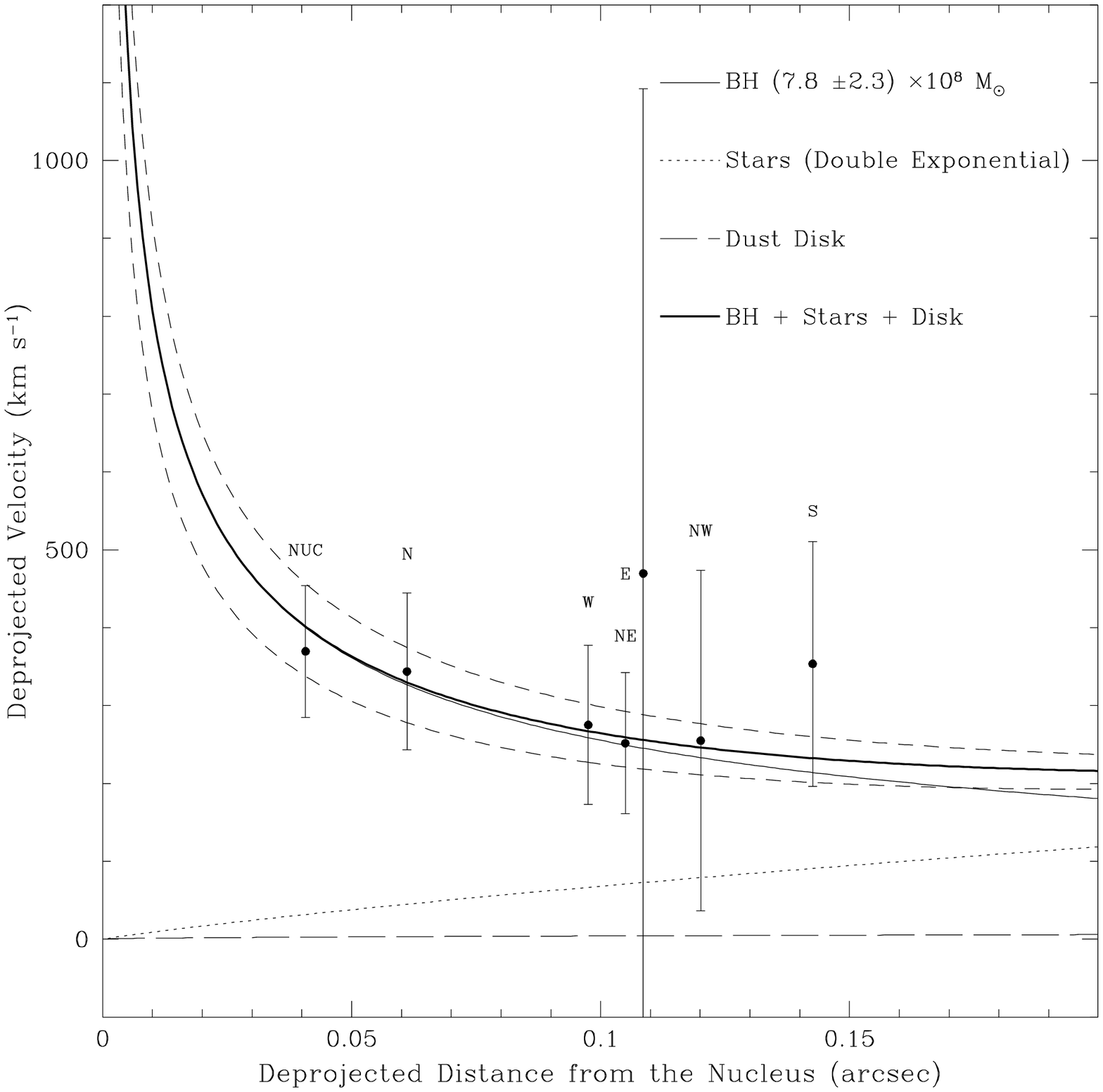}
\caption{Comparison between the measured velocities and the best fitting BH
model derived by using a double exponential profile to describe the
stellar surface brightness. The dotted line shows the expected rotation
curve due to the stars for $(M/L)_V = 8.5$ \sm/\sl. The long$-$dashed line shows
the contribution to the velocity due to the dust disk. The thin solid
line is the Keplerian rotation due to a $7.8 \times 10^8$ M$_\odot$
black hole, while the thick solid line shows the final velocity taking
into account the stars, the disk,  and the black hole. The dashed lines
represent the 1$\sigma$ limit on the final velocity due to the
uncertainty in the black hole mass ($\pm 2.3 M_{\odot}$).} 
\end{figure}

\clearpage

\begin{figure}
\figurenum{15}
\plotone{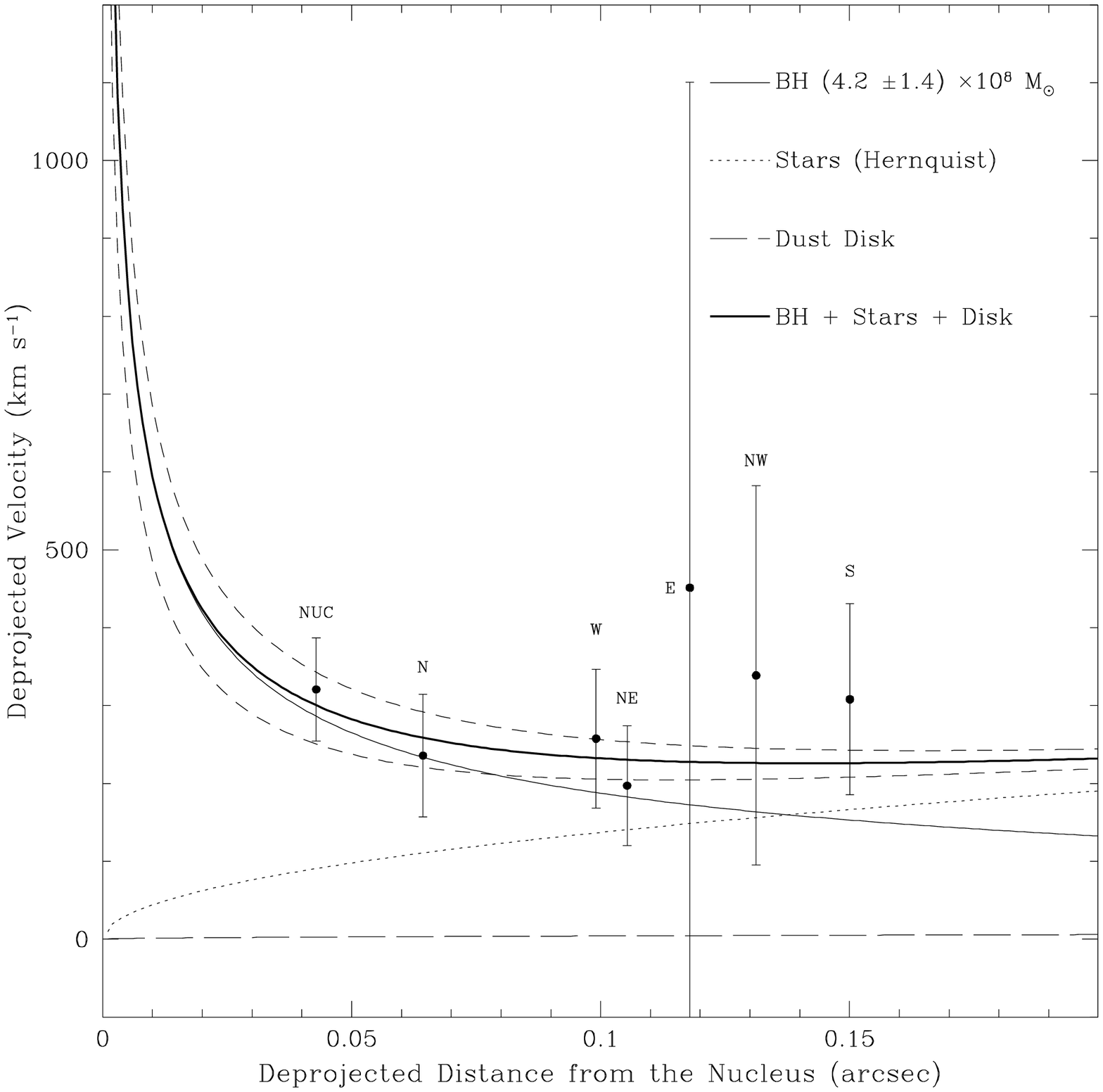}
\caption{Comparison between the measured velocities and the best fitting BH
model derived by using the Hernquist model to describe the
stellar surface brightness. The dotted line shows the expected rotation
curve due to the stars for $(M/L)_V = 8.5$ \sm/\sl. The long$-$dashed line shows
the contribution to the velocity due to the dust disk. The thin solid
line is the Keplerian rotation due to a $4.2 \times 10^8$ M$_\odot$
black hole, while the thick solid line shows the final velocity taking
into account the stars, the disk,  and the black hole. The dashed lines
represent the 1$\sigma$ limit on the final velocity due to the
uncertainty in the black hole mass ($\pm 1.4 M_{\odot}$).} 
\end{figure}

\clearpage

\begin{figure}
\figurenum{16}
\plotone{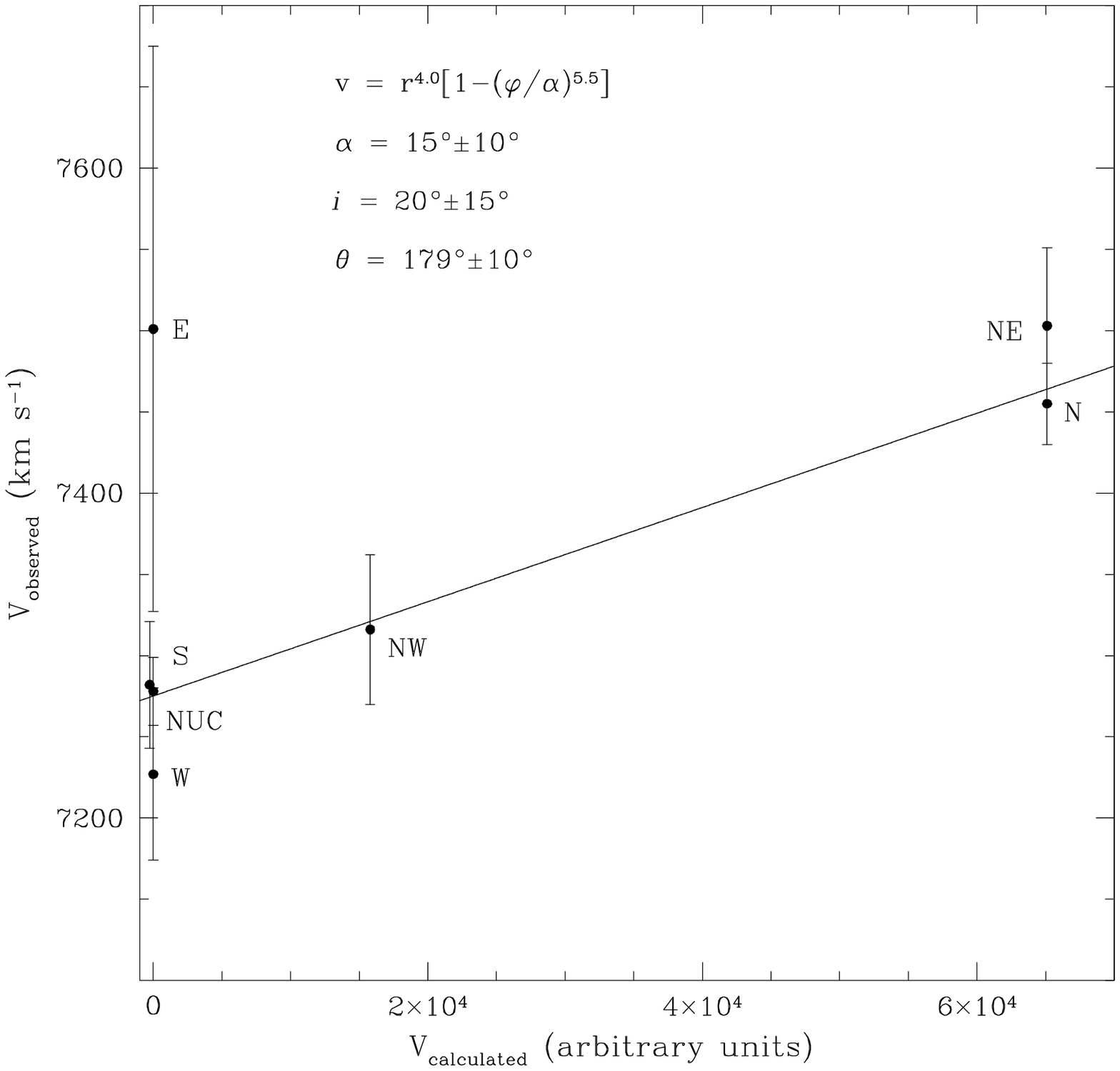}
\caption{The velocities predicted by the best fitting biconical outflow model (\S 4.2) plotted against the measured velocities.
The parameters of the best fit model are summarized at the top left
corner of the Figure and are discussed in the text (\S 5.1).}
\end{figure}

\clearpage

\begin{figure}
\figurenum{17}
\plotone{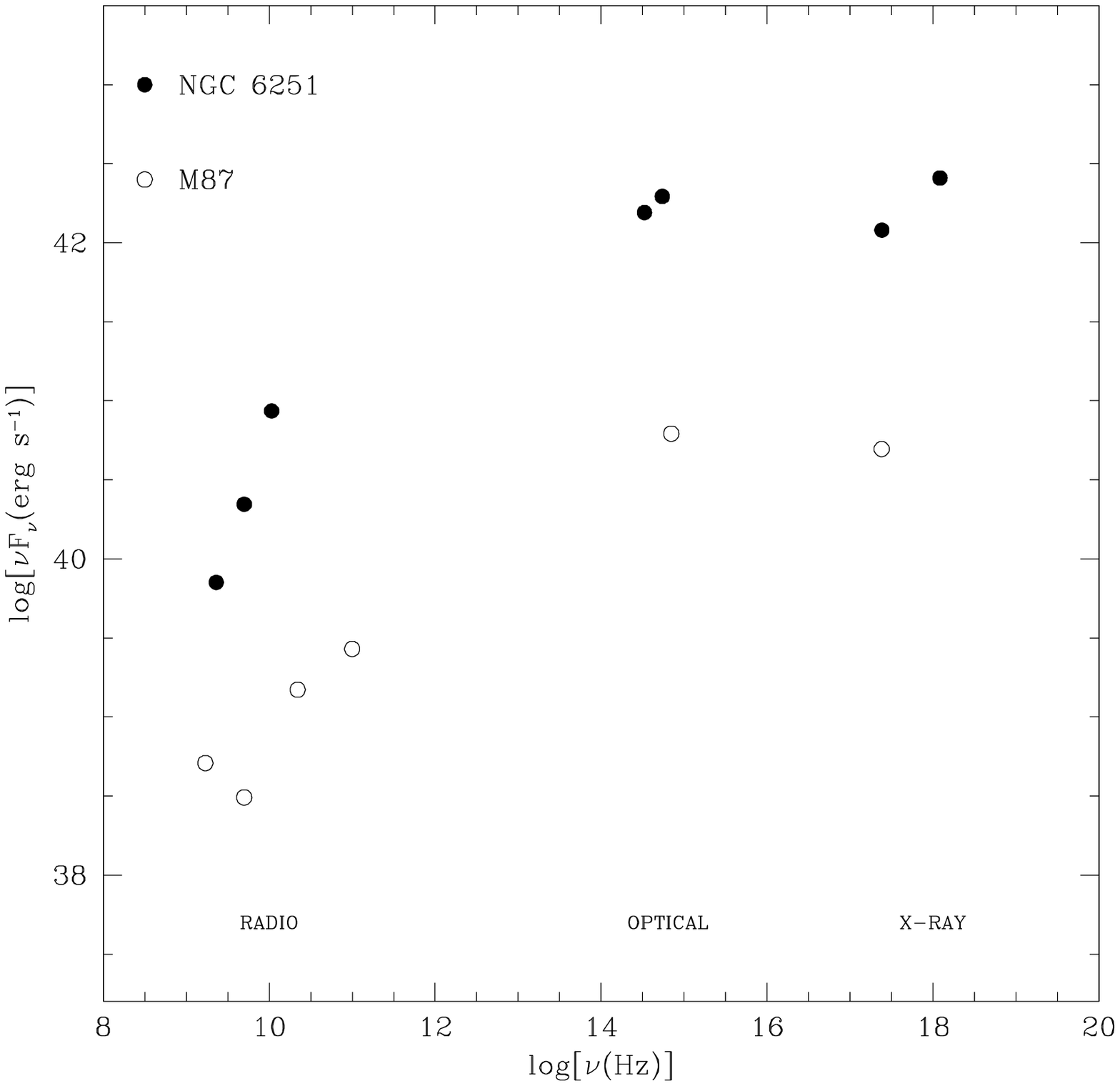}
\caption{Radio, optical and X$-$ray measurements of the nuclear non$-$thermal luminosity for NGC 6251 and M87 (references in the text).}
\end{figure}

\clearpage

\begin{deluxetable}{llll}
\tablecolumns{9} \tablewidth{0pc} \scriptsize
\tablecaption{Astrometry\label{tbl-1}} \tablehead{ \colhead{} &
\colhead{RA (h m s, J2000)} & \colhead{Dec
($^\circ$~$^\prime$~$^{\prime\prime}$, J2000)} &
\colhead{PA(1996.7)}\tablenotemark{a}\\ } \startdata Nucleus
$V$\tablenotemark{b}   & 16 32 32.154 $\pm$ 0.001 & 82 32 16.563 $\pm$
0.009 &\nodata\nl Nucleus $I$\tablenotemark{b}   & 16 32 32.158 $\pm$
0.001 & 82 32 16.590 $\pm$ 0.009 &\nodata\nl Isophotes
$V$\tablenotemark{b,c} & 16 32 32.154 $\pm$ 0.001 & 82 32 16.549 $\pm$
0.009& 20\Deg7 $\pm$ 1\Deg4\nl Isophotes $I$\tablenotemark{b,c} & 16
32 32.160 $\pm$ 0.001 & 82 32 16.580 $\pm$ 0.009& 20\Deg8 $\pm$
1\Deg0\nl Dust Disk $I$\tablenotemark{b}      & 16 32 32.16  $\pm$
0.003  & 82 32 16.68  $\pm$ 0.04 & 4\deg $\pm$ 2\deg\nl Dust Disk
$V$\tablenotemark{b}      & 16 32 32.16  $\pm$ 0.003  & 82 32 16.66
$\pm$ 0.04 & 4\deg $\pm$ 2\deg\nl Inner Gas Disk&  16 32 32.153 $\pm$
0.001 & 82 32 16.567 $\pm$ 0.018 & 59\deg $\pm$ 10\deg\nl & & &\nl
1665 MHz radio core\tablenotemark{d}   & 16 32 31.77 & 82 32 16.16
&\nodata\nl Radio Jet Axis\tablenotemark{e}~($< 10$mas) & \nodata &
\nodata & 297\Deg1 $\pm$ 1\Deg4\nl Radio Jet Axis\tablenotemark{e}~($<
25$mas) & \nodata & \nodata & 294\Deg3 $\pm$ 1\Deg0\nl Radio Jet
Axis\tablenotemark{e}~(Kpc scale) & \nodata & \nodata & 293\Deg8 $\pm$
0\Deg1\nl & & &\nl Stellar Rotation Axis\tablenotemark{f}
&\nodata  & \nodata & 85\deg   $\pm$ 28\deg\nl \enddata
\tablenotetext{a}{All position angles have be precessed to the epoch
of the WFPC2 observations, 1996.7} 
\tablenotetext{b}{All coordinates
derived from WFPC2 measurements can suffer from a systematic error of
up to 1\sec, due to errors in the guide stars positions and internal
instrumental effects, such as the filter wedge effect (Biretta et
al. 1996). Note that because of the filter wedge effect, relative
positions of objects measured in different filters (for example the position of
the nucleus in V and in I) are accurate only to within a few
pixels. However, relative positions within the same image (for example
the position of the nucleus and of the center of the disk in the V
image) are accurate to within the quoted uncertainty}
\tablenotetext{c}{The isophotal center is averaged between 1\Sec5 and
4\Sec0.}  \tablenotetext{d}{From Jones \& Wehrle (1994)}
\tablenotetext{e}{From Jones et al. (1986)} \tablenotetext{f}{From
Heckman et al. (1985)} 
\end{deluxetable}

\begin{deluxetable}{clccccccc}
\tablecolumns{9}
\tablewidth{0pc}
\scriptsize
\tablecaption{Line Fitting Parameters\label{tbl-2}}
\tablehead{
\colhead{Line} &
\colhead{Parameter} &
\multicolumn{7}{c} {Aperture}\\
\cline{3-9}\\
\colhead{} & \colhead{} & \colhead{NUC} & \colhead{N} & \colhead{S} & \colhead{W}
 & \colhead{E} & \colhead{NW} & \colhead{NE} 
}
\startdata
[OI]\dl6300& v\tablenotemark{a}     &7270 $\pm$   32  &  7467 $\pm$   63  &  7288 $\pm$ 252  &  7227 $\pm$  295  &  7501 $\pm$ 321  &  7310 $\pm$  382  &  7509 $\pm$  290\nl
"& FWHM\tablenotemark{a}    &2000 $\pm$  101  &  1809 $\pm$  145  &  1847 $\pm$ 513  &  1400 $\pm$  409  &  2962 $\pm$ 618  &  2800 $\pm$  581  &  1526 $\pm$  223\nl
"& Flux\tablenotemark{b} &3.48 $\pm$ 0.21  &  2.46 $\pm$ 0.15  &  0.33 $\pm$ 0.05  &  0.40 $\pm$ 0.07  &  0.51 $\pm$ 0.38  &  0.60 $\pm$ 0.11  &  0.81 $\pm$ 0.11\nl
 & & & & & & & & \nl
[OI]\dl6364& Flux\tablenotemark{b} &1.20 $\pm$ 0.14  &  1.00 $\pm$ 0.11  &  0.04 $\pm$ 0.06  &  0.25 $\pm$ 0.07  &  0.39 $\pm$ 0.11  &  0.40 $\pm$ 0.12  &  0.20 $\pm$ 0.08\nl
 & & & & & & & & \nl
[OII]\dl7325\tablenotemark{c}& v\tablenotemark{a}     &7200 $\pm$  131  &  \nodata &  \nodata &  \nodata &  \nodata &  \nodata &  \nodata \nl
"& FWHM\tablenotemark{a}    &2039 $\pm$ 397 &  \nodata &  \nodata &  \nodata &  \nodata &  \nodata &  \nodata \nl
"& Flux\tablenotemark{b} &0.97 $\pm$ 0.15 &  \nodata &  \nodata &  \nodata &  \nodata &  \nodata &  \nodata \nl
 & & & & & & & & \nl
[NII]\dl6584& v\tablenotemark{a}     &7278 $\pm$   21  &  7455 $\pm$   25  &  7282 $\pm$   39  &  7227 $\pm$   53  &  7501 $\pm$  174  &  7316 $\pm$   46  &  7503 $\pm$   48\nl
"& FWHM\tablenotemark{a}    &1021 $\pm$   78  &  1077 $\pm$   91  &   928 $\pm$  137  &  1103 $\pm$  208  &  1069 $\pm$  262  &   850 $\pm$  114  &  1236 $\pm$  130\nl
"& Flux\tablenotemark{b} &5.74 $\pm$ 0.28  &  5.36 $\pm$ 0.32  &  0.79 $\pm$ 0.08  &  0.84 $\pm$ 0.11  &  0.88 $\pm$ 0.21  &  0.80 $\pm$ 0.10  &  1.82 $\pm$ 0.21\nl
"& Skew\tablenotemark{d} & 0.98 $\pm$ 0.12 & 0.93 $\pm$ 0.12 &  \nodata &  \nodata &  \nodata &  \nodata &  \nodata \nl
 & & & & & & & & \nl
\ha~narrow& Flux\tablenotemark{b} &1.53 $\pm$ 0.22  &  1.20 $\pm$ 0.22  &  0.41 $\pm$ 0.07  &  0.16 $\pm$ 0.10  &  0.30 $\pm$ 0.13  &  0.18 $\pm$ 0.07  &  0.36 $\pm$ 0.16\nl
 & & & & & & & & \nl
\ha~broad\tablenotemark{e}& v\tablenotemark{a}     &6392 $\pm$    356  &   6450 $\pm$ 420 &  \nodata &  \nodata &  \nodata &  \nodata &  \nodata \nl
"& FWHM\tablenotemark{a}    &3656 $\pm$   450  &  3450 $\pm$ 561 &  \nodata &  \nodata &  \nodata &  \nodata &  \nodata \nl
"& Flux\tablenotemark{b} &6.31 $\pm$ 0.47  &  4.03 $\pm$ 0.36  &  0.60 $\pm$ 0.40  &  0.52 $\pm$ 0.40  &  0.50 $\pm$ 0.32  &  0.83 $\pm$ 0.19  &  1.21 $\pm$ 0.30\nl
"& Skew & 2.2 $\pm$ 0.5 & 2.1 $\pm$ 0.3 &  \nodata &  \nodata &  \nodata &  \nodata &  \nodata \nl
 & & & & & & & & \nl
[SII]\dl6717& v\tablenotemark{a}     &7281 $\pm$  312  &  7450 $\pm$  378  &  7280 $\pm$  412  &  7284 $\pm$ 450  &  7503 $\pm$  395  &  7314 $\pm$ 432  &  7516 $\pm$  441\nl
"& FWHM\tablenotemark{a}    &1399 $\pm$  224  &  1159 $\pm$  355  &   500 $\pm$  357  &   800 $\pm$  376  &   871 $\pm$  327  &   800 $\pm$  353  &   800 $\pm$  374\nl
"& Flux\tablenotemark{b} &0.92 $\pm$ 2.02  &  0.84 $\pm$ 1.07  &  0.10 $\pm$ 0.04  &  0.20 $\pm$ 0.03  &  0.23 $\pm$ 0.10  &  0.18 $\pm$ 0.05  &  0.35 $\pm$ 0.09\nl
 & & & & & & & & \nl
[SII]\dl6731& Flux\tablenotemark{b} &1.25 $\pm$ 2.08  &  1.23 $\pm$ 1.16  &  0.14 $\pm$ 0.04  &  0.20 $\pm$ 0.03  &  0.11 $\pm$ 0.09  &  0.12 $\pm$ 0.02  &  0.35 $\pm$ 0.12\nl
\enddata
\tablenotetext{a}{Velocity and FWHM are given in \kms.
The redshift and FWHM of [OI]\dl6364, [NII]\dl6548 and [SII]\dl6731
are constrained to be the same as those for [OI]\dl6300, [NII]\dl6584 and [SII]\dl6717
respectively. Also the redshift and FWHM of the narrow \ha~line is the same as for the NII lines.}
\tablenotetext{b}{Fluxes are given in counts s\ma. The ratio of the fluxes of [NII]\dl6584
and [NII]\dl6548 is constrained to be equal to 3.0.}
\tablenotetext{c}{[OII]\dl7325 is not detected at positions other than NUC.}
\tablenotetext{d}{The skew parameter is constrained to be the same for all narrow emission lines. In addition,
at positions S, W, E, NW and NE the skew parameter is set equal to the value measured at NUC, 0.98.}
\tablenotetext{e}{The central wavelength, FWHM and skew parameter for the broad \ha~component at positions S, W, E, NW and NE are set equal to the nuclear values.}
\end{deluxetable}

\begin{deluxetable}{lll}
\tablecolumns{3} \tablewidth{0pc} \scriptsize
\tablecaption{Predictions of the Keplerian Disk Model \label{tbl-3}} 
\tablehead{ 
\colhead{ } &
\colhead{Double Exponential} &
\colhead{Hernquist}\\ 
\colhead{ } &
\colhead {Stellar Model} &
\colhead {Stellar Model} \\ }
\startdata 
Inclination\tablenotemark{a}       & 32\deg $\pm$ 15\deg        & 39\deg $\pm$ 15\deg        \nl       
Position Angle\tablenotemark{b}    & 62\deg $\pm$ 10\deg        & 57\deg $\pm$ 10\deg        \nl
$\Delta$RA\tablenotemark{c}        & 0\Sec00 $\pm$ 0\Sec018     & 0\Sec00 $\pm$ 0\Sec018     \nl
$\Delta$Dec\tablenotemark{c}        & 0\Sec054 $\pm$ 0\Sec018    & 0\Sec054 $\pm$ 0\Sec018    \nl
$M_{BH}$\tablenotemark{d}          & $(7.8 \pm 2.3)\times10^8$  & $(4.2 \pm 1.4)\times10^8$  \nl
Systemic Velocity\tablenotemark{e} & 7370 $\pm$ 13              & 7368 $\pm$ 13              \nl
\enddata
\tablenotetext{a}{Predicted angle between the normal to the disk and the line of sight (\S 4.1).} 
\tablenotetext{b}{Predicted position angle of the major axis of the disk (\S 4.1).}
\tablenotetext{c}{Predicted displacements, in RA and Dec, of the central FOS aperture with respect to the center of the disk (\S 4.1).}  
\tablenotetext{d}{Predicted central mass concentration (in \sm, \S 4.1).}
\tablenotetext{e}{Predicted systemic velocity (in \kms, \S 4.1). The ZCAT systemic velocity is 7400 $\pm$ 22 \kms.} 
\end{deluxetable}

\end{document}